\documentclass[aps,prd,notitlepage,amsmath,amssymb,preprintnumbers,nofootinbib,longbibliography,superscriptaddress,10pt]{revtex4-1}
\usepackage[utf8]{inputenc} 
\usepackage{graphicx}
\usepackage{url}
\usepackage[bookmarks, pagebackref=false]{hyperref}
\usepackage[usenames,dvipsnames]{xcolor}
\usepackage{todonotes}
\usepackage{tikz}
\usetikzlibrary{shapes,decorations.pathreplacing,decorations.markings,positioning}
\allowdisplaybreaks
\makeatletter
\pgfdeclareshape{crossed circle}
{
	\inheritsavedanchors[from=circle]
	\inheritanchorborder[from=circle]
	\inheritanchor[from=circle]{north}
	\inheritanchor[from=circle]{north west}
	\inheritanchor[from=circle]{north east}
	\inheritanchor[from=circle]{center}
	\inheritanchor[from=circle]{west}
	\inheritanchor[from=circle]{east}
	\inheritanchor[from=circle]{mid}
	\inheritanchor[from=circle]{mid west}
	\inheritanchor[from=circle]{mid east}
	\inheritanchor[from=circle]{base}
	\inheritanchor[from=circle]{base west}
	\inheritanchor[from=circle]{base east}
	\inheritanchor[from=circle]{south}
	\inheritanchor[from=circle]{south west}
	\inheritanchor[from=circle]{south east}
	
	\inheritbackgroundpath[from=circle]
	
	\foregroundpath{
		\centerpoint
		\pgf@xc =\radius
		\advance\pgf@x by-0.707107\pgf@xc
		\advance\pgf@y by-0.707107\pgf@xc
		\pgf@xa=\pgf@x \pgf@ya=\pgf@y  
		\centerpoint
		\pgf@xc=\radius
		\advance\pgf@x by0.707107\pgf@xc
		\advance\pgf@y by0.707107\pgf@xc
		\pgf@xb=\pgf@x \pgf@yb=\pgf@y  
		\pgfpathmoveto{\pgfqpoint{\pgf@xa}{\pgf@ya}}
		\pgfpathlineto{\pgfqpoint{\pgf@xb}{\pgf@yb}}
		\pgfpathmoveto{\pgfqpoint{\pgf@xa}{\pgf@yb}}
		\pgfpathlineto{\pgfqpoint{\pgf@xb}{\pgf@ya}}
		\pgfsetarrowsstart{}
		\pgfsetarrowsend{}
	}
}
\pgfdeclareshape{crossed rectangle}
{
	\inheritsavedanchors[from=rectangle]
	\inheritanchorborder[from=rectangle]
	\inheritanchor[from=rectangle]{north}
	\inheritanchor[from=rectangle]{north west}
	\inheritanchor[from=rectangle]{north east}
	\inheritanchor[from=rectangle]{center}
	\inheritanchor[from=rectangle]{west}
	\inheritanchor[from=rectangle]{east}
	\inheritanchor[from=rectangle]{mid}
	\inheritanchor[from=rectangle]{mid west}
	\inheritanchor[from=rectangle]{mid east}
	\inheritanchor[from=rectangle]{base}
	\inheritanchor[from=rectangle]{base west}
	\inheritanchor[from=rectangle]{base east}
	\inheritanchor[from=rectangle]{south}
	\inheritanchor[from=rectangle]{south west}
	\inheritanchor[from=rectangle]{south east}
	
	\inheritbackgroundpath[from=rectangle]
	
	\foregroundpath{
		\southwest \pgf@xa=\pgf@x \pgf@ya=\pgf@y
		\northeast \pgf@xb=\pgf@x \pgf@yb=\pgf@y
		\pgfpathmoveto{\pgfqpoint{\pgf@xa}{\pgf@ya}}
		\pgfpathlineto{\pgfqpoint{\pgf@xb}{\pgf@yb}}
		\pgfpathmoveto{\pgfqpoint{\pgf@xa}{\pgf@yb}}
		\pgfpathlineto{\pgfqpoint{\pgf@xb}{\pgf@ya}}
		\pgfsetarrowsstart{}
		\pgfsetarrowsend{}
	}
}
\pgfdeclarelayer{edgelayer}
\pgfdeclarelayer{nodelayer}
\pgfsetlayers{edgelayer,nodelayer,main}
\tikzset{
	PL/.style={myred,thick},
	PR/.style={myblue,thick},
	PLL/.style={black,thick}
}
\tikzset{
	none/.style={inner sep=0pt},
	h4/.style = {
		minimum size=12pt,
		fill={rgb,255: red,128; green,128; blue,0}, 
		draw={rgb,255: red,128; green,128; blue,0}, 
		shape=circle, 
		scale=0.3
	},
	ffh/.style={
		fill=mycyan,
		shape=circle,
		scale=0.3
	},
	Op/.style={
		draw,
		fill=white,
		shape=crossed circle,
		scale=0.5
	},
	ChFlip/.style={
		draw,
		fill=white,
		shape=crossed rectangle,
		scale=0.5
	},
	onepi/.style={
		circle,
		minimum height=15pt,
		minimum width=15pt,
		fill=gray
	},
	scalar/.style={
		dashed,
		draw={rgb,255: red,255; green,128; blue,0}
	},
	sc/.style={
		thick,
		draw={black}
	},
	fermion/.style={
		postaction=decorate,
		decoration={
			markings,
			mark=at position #1 with {
				\node[transform shape,
				isosceles triangle,
				inner sep=0mm,
				minimum width=3pt,
				xshift=-0.5pt,
				draw=none,
				fill] {};
			},
		}
	},
	fermion/.default={0.5},
	HtoHbar/.style n args={3}{
		preaction=decorate,decoration={
			markings,
			mark=at position #1 with {
				\draw[stealth-,thin,#2] (-1.5mm,1.5mm) -- (1.5mm,1.5mm);
				\draw[-stealth,thin,#3] (-1.5mm,-1.5mm) -- (1.5mm,-1.5mm);
			}
		}
	},
	HtoHbar/.default={0.4}{PL}{PR},
	HtoHbar'/.style n args={3}{
		preaction=decorate,decoration={
			markings,
			mark=at position #1 with {
				\draw[-stealth,thin,#2] (-1.5mm,1.5mm) -- (1.5mm,1.5mm);
				\draw[stealth-,thin,#3] (-1.5mm,-1.5mm) -- (1.5mm,-1.5mm);
			}
		}
	},
	HtoHbar'/.default={0.4}{PR}{PL}
}
\tikzset{every picture/.style={baseline=+12pt,scale=0.5}}

\tikzset{
	s1/.pic={
		\begin{pgfonlayer}{nodelayer}
			\node [style=ChFlip] (0) at (-6.25, 1.5) {};
			\node [style=ChFlip] (1) at (-3.75, 1.5) {};
			\node [style=ChFlip] (2) at (-1.25, 1.5) {};
			\node [style=ChFlip] (3) at (-3.75, -1) {};
			\node [style=none,minimum size=4pt,label={[xshift=-1pt,yshift=-4pt]$a$}] (a) at (-7.75, 2.5) {};
			\node [style=none,minimum size=4pt,label={[xshift=1pt,yshift=-4pt]$b$}] (b) at (-5, 2.5) {};
			\node [style=none,minimum size=4pt,label={[xshift=2pt,yshift=-4pt]$c$}] (c) at (-2.5, 2.5) {};
			\node [style=none,minimum size=4pt,label={[xshift=3pt,yshift=-4pt]$d$}] (d) at (0.2, 2.5) {};
			\node [style=none,minimum size=4pt,label={[xshift=-1pt,yshift=-4pt]$$}] (ap) at (-7.25, 2.25) {};
			\node [style=none,minimum size=4pt,label={[xshift=1pt,yshift=-4pt]$$}] (bp) at (-4.75, 2.25) {};
			\node [style=none,minimum size=4pt,label={[xshift=2pt,yshift=-4pt]$$}] (cp) at (-2.25, 2.25) {};
			\node [style=none,minimum size=4pt,label={[xshift=3pt,yshift=-4pt]$$}] (dp) at (-0.25, 2.25) {};
		\end{pgfonlayer}
		\begin{pgfonlayer}{edgelayer}
			\draw [style={fermion},PL] (b.south) to [out=15,in=-135] (0.east);
			\draw [style={fermion},PR] (0.west) to [out=-45,in=175] (ap.south) ;
			\draw [style={fermion},PL] (c.south) to[ out=15,in=-135] (1.east) ;
			\draw [style={fermion},PR](1.west) to [out=-45,in=175]  (bp.south) ;
			\draw [style={fermion},PL] (dp.south) to[out=15,in=-135] (2.east);
			\draw [style={fermion},PR] (2.west) to[out=-45,in=175] (cp.south) ;
			\draw [style={fermion},PL] (a.south) to[out=-65,in=195]  (3.west) ;
			\draw [style={fermion},PR]  (3.east) to[out=0,in=-105] (d.south);
			
		\end{pgfonlayer}
	},
	s2/.pic={
		\begin{pgfonlayer}{nodelayer}
			\node [style=ChFlip] (0) at (-6.25, 1.5) {};
			\node [style=ChFlip] (2) at (-1.25, 1.5) {};
			\node [style=none,minimum size=4pt,label={[xshift=-1pt,yshift=-4pt]$a$}] (a) at (-7.5, 2.5) {};
			\node [style=none,minimum size=4pt,label={[xshift=1pt,yshift=-4pt]$b$}] (b) at (-5, 2.5) {};
			\node [style=none,minimum size=4pt,label={[xshift=2pt,yshift=-4pt]$c$}] (c) at (-2.5, 2.5) {};
			\node [style=none,minimum size=4pt,label={[xshift=3pt,yshift=-4pt]$d$}] (d) at (0, 2.5) {};
		\end{pgfonlayer}
		\begin{pgfonlayer}{edgelayer}
			\draw [style={fermion},PL] (b.south)  to[out=15,in=-135] (0.east) ;
			\draw [style={fermion},PR]  (0.west) to[out=-45,in=175]  (a.south);
			\draw [style={fermion},PR] (c.west) to[out=-135,in=-45] (b.east);
			\draw [style={fermion},PR] (d.south)  to[out=15,in=-135] (2.east);
			\draw [style={fermion},PL] (2.west) to[out=-45,in=175] (c.south) ;
			\draw [style={fermion},PL] (a.west) to[out=-135,in=-45] (d.east);
		\end{pgfonlayer}
	},
	s3/.pic={
		\begin{pgfonlayer}{nodelayer}
			\node [style=ChFlip] (0) at (-6.25, 1.5) {};
			\node [style=ChFlip] (1) at (-3.75, 1.5) {};
			\node [style=none,minimum size=4pt,label={[xshift=-1pt,yshift=-4pt]$a$}] (a) at (-7.5, 2.5) {};
			\node [style=none,minimum size=4pt,label={[xshift=1pt,yshift=-4pt]$b$}] (b) at (-5, 2.5) {};
			\node [style=none,minimum size=4pt,label={[xshift=1pt,yshift=-4pt]$$}] (bp) at (-4.75, 2.25) {};
			\node [style=none,minimum size=4pt,label={[xshift=2pt,yshift=-4pt]$c$}] (c) at (-2.5, 2.5) {};
			\node [style=none,minimum size=4pt,label={[xshift=3pt,yshift=-4pt]$d$}] (d) at (0, 2.5) {};
		\end{pgfonlayer}
		\begin{pgfonlayer}{edgelayer}
			\draw [style={fermion},PL] (b.south) to[out=15,in=-135] (0.east);
			\draw [style={fermion},PR] (0.west) to[out=-45,in=175]  (a.south);
			\draw [style={fermion},PL]  (c.south) to[out=15,in=-135](1.east);
			\draw [style={fermion},PR] (1.west) to[out=-45,in=175]  (bp.south);
			\draw [style={fermion},PR] (d.west) to[out=-135,in=-45] (c.east);
			\draw [style={fermion},PL] (a.west) to[out=-135,in=-45] (d.east);
		\end{pgfonlayer}
	},
	d2/.pic={
		\begin{pgfonlayer}{nodelayer}
			\node [style=ChFlip] (0) at (-6.25, 1.5) {};
			\node [style=ChFlip] (1) at (-6.25, 2.5) {};
			\node [style=ChFlip] (2) at (-1.25, 1.5) {};
			\node [style=ChFlip] (3) at (-1.25, 2.5) {};
			\node [style=none,minimum size=4pt,label={[xshift=-1pt,yshift=-4pt]$a$}] (a) at (-7.5, 2.5) {};
			\node [style=none,minimum size=4pt,label={[xshift=1pt,yshift=-4pt]$b$}] (b) at (-5, 2.5) {};
			\node [style=none,minimum size=4pt,label={[xshift=2pt,yshift=-4pt]$c$}] (c) at (-2.5, 2.5) {};
			\node [style=none,minimum size=4pt,label={[xshift=3pt,yshift=-4pt]$d$}] (d) at (0, 2.5) {};
			\node [style=none,minimum size=4pt,label={[xshift=-1pt,yshift=-4pt]$$}] (ap) at (-7.25, 2.25) {};
			\node [style=none,minimum size=4pt,label={[xshift=1pt,yshift=-4pt]$$}] (bp) at (-4.75, 2.25) {};
			\node [style=none,minimum size=4pt,label={[xshift=2pt,yshift=-4pt]$$}] (cp) at (-2.25, 2.25) {};
			\node [style=none,minimum size=4pt,label={[xshift=3pt,yshift=-4pt]$$}] (dp) at (-0.5, 2.25) {};
		\end{pgfonlayer}
		\begin{pgfonlayer}{edgelayer}
			
			\draw [style={fermion},PL] (bp.south)  to[out=15,in=-135] (0.east) ;
			\draw [style={fermion},PR]  (0.west) to[out=-45,in=175]  (ap.south);
			\draw [style={fermion},PR]  (1.east) to[out=15,in=-135] (b.south) ;
			\draw [style={fermion},PL]  (a.south) to[out=-45,in=175]  (1.west);

			\draw [style={fermion},PL] (dp.south)  to[out=15,in=-135] (2.east) ;
			\draw [style={fermion},PR]  (2.west) to[out=-45,in=175]  (cp.south);
			\draw [style={fermion},PR]  (3.east) to[out=15,in=-135] (d.south) ;
			\draw [style={fermion},PL]  (c.south) to[out=-45,in=175]  (3.west);
		\end{pgfonlayer}
	},
	d1/.pic={
		\begin{pgfonlayer}{nodelayer}
			\node [style=ChFlip] (0) at (-6.25, 1.5) {};
			\node [style=ChFlip] (1) at (-6.25, 2.5) {};
			\node [style=ChFlip] (2) at (-1.25, 1.5) {};
			\node [style=ChFlip] (3) at (-1.25, 2.5) {};
			\node [style=none,minimum size=4pt,label={[xshift=-1pt,yshift=-4pt]$a$}] (a) at (-7.5, 2.5) {};
			\node [style=none,minimum size=4pt,label={[xshift=1pt,yshift=-4pt]$b$}] (b) at (-5, 2.5) {};
			\node [style=none,minimum size=4pt,label={[xshift=2pt,yshift=-4pt]$c$}] (c) at (-2.5, 2.5) {};
			\node [style=none,minimum size=4pt,label={[xshift=3pt,yshift=-4pt]$d$}] (d) at (0, 2.5) {};
			\node [style=none,minimum size=4pt,label={[xshift=-1pt,yshift=-4pt]$$}] (ap) at (-7.25, 2.25) {};
			\node [style=none,minimum size=4pt,label={[xshift=1pt,yshift=-4pt]$$}] (bp) at (-4.75, 2.25) {};
			\node [style=none,minimum size=4pt,label={[xshift=2pt,yshift=-4pt]$$}] (cp) at (-2.25, 2.25) {};
			\node [style=none,minimum size=4pt,label={[xshift=3pt,yshift=-4pt]$$}] (dp) at (-0.5, 2.25) {};
		\end{pgfonlayer}
		\begin{pgfonlayer}{edgelayer}
			
			\draw [style={fermion},PL] (bp.south)  to[out=15,in=-135] (0.east) ;
			\draw [style={fermion},PR]  (0.west) to[out=-45,in=175]  (ap.south);
			\draw [style={fermion},PR]  (1.east) to[out=15,in=-135] (b.south) ;
			\draw [style={fermion},PL]  (a.south) to[out=-45,in=175]  (1.west);

			\draw [style={fermion},PR] (dp.south)  to[out=15,in=-135] (2.east) ;
			\draw [style={fermion},PL]  (2.west) to[out=-45,in=175]  (cp.south);
			\draw [style={fermion},PL]  (3.east) to[out=15,in=-135] (d.south) ;
			\draw [style={fermion},PR]  (c.south) to[out=-45,in=175]  (3.west);
		\end{pgfonlayer}
	},
	d3/.pic={
		\begin{pgfonlayer}{nodelayer}
			\node [style=ChFlip] (0) at (-6.25, 1.5) {};
			\node [style=ChFlip] (1) at (-6.25, 2.5) {};
			\node [style=none,minimum size=4pt,label={[xshift=-1pt,yshift=-4pt]$a$}] (a) at (-7.5, 2.5) {};
			\node [style=none,minimum size=4pt,label={[xshift=1pt,yshift=-4pt]$b$}] (b) at (-5, 2.5) {};
			\node [style=none,minimum size=4pt,label={[xshift=2pt,yshift=-4pt]$c$}] (c) at (-2.5, 2.5) {};
			\node [style=none,minimum size=4pt,label={[xshift=3pt,yshift=-4pt]$d$}] (d) at (0, 2.5) {};
			\node [style=none,minimum size=4pt,label={[xshift=-1pt,yshift=-4pt]$$}] (ap) at (-7.25, 2.25) {};
			\node [style=none,minimum size=4pt,label={[xshift=1pt,yshift=-4pt]$$}] (bp) at (-4.75, 2.25) {};
		\end{pgfonlayer}
		\begin{pgfonlayer}{edgelayer}
			
			\draw [style={fermion},PL] (bp.south)  to[out=15,in=-135] (0.east) ;
			\draw [style={fermion},PR]  (0.west) to[out=-45,in=175]  (ap.south);
			\draw [style={fermion},PR]  (1.east) to[out=15,in=-135] (b.south) ;
			\draw [style={fermion},PL]  (a.south) to[out=-45,in=175]  (1.west);

			\draw [style={fermion},PL] (c.south) to[out=-45,in=-135] (d.south);
			\draw [style={fermion},PR] (d.west) to[out=-135,in=-45] (c.east);
		\end{pgfonlayer}
	},
	d4/.pic={
		\begin{pgfonlayer}{nodelayer}
			\node [style=ChFlip] (0) at (-6.25, 1.5) {};
			\node [style=ChFlip] (1) at (-3.75, 1.5) {};
			\node [style=ChFlip] (2) at (-4.35, 0) {};
			\node [style=ChFlip] (3) at (-1, 1.5) {};
			\node [style=none,minimum size=4pt,label={[xshift=-1pt,yshift=-4pt]$a$}] (a) at (-7.85, 2.5) {};
			\node [style=none,minimum size=4pt,label={[xshift=1pt,yshift=-4pt]$b$}] (b) at (-5, 2.5) {};
			\node [style=none,minimum size=4pt,label={[xshift=2pt,yshift=-4pt]$c$}] (c) at (-2.5, 2.5) {};
			\node [style=none,minimum size=4pt,label={[xshift=3pt,yshift=-4pt]$d$}] (d) at (0, 2.5) {};
			\node [style=none,minimum size=4pt,label={[xshift=-1pt,yshift=-4pt]$$}] (ap) at (-7.25, 2.25) {};
			\node [style=none,minimum size=4pt,label={[xshift=1pt,yshift=-4pt]$$}] (bp) at (-4.75, 2.25) {};
			\node [style=none,minimum size=4pt,label={[xshift=2pt,yshift=-4pt]$$}] (cp) at (-2.25, 2.25) {};
		\end{pgfonlayer}
		\begin{pgfonlayer}{edgelayer}
			\draw [style={fermion},PL] (b.south) to [out=15,in=-135] (0.east);
			\draw [style={fermion},PR] (0.west) to [out=-45,in=175] (ap.south) ;
			\draw [style={fermion},PL] (c.south) to[ out=15,in=-135] (1.east) ;
			\draw [style={fermion},PR](1.west) to [out=-45,in=175]  (bp.south) ;
			\draw [style={fermion},PR] (2.east) to[out=15,in=-135] (cp.south);
			\draw [style={fermion},PL] (a.south) to[out=-45,in=175] (2.west)  ;
			\draw [style={fermion},PR] (3.east) to[out=15,in=-135] (d.south);
			\draw [style={fermion},PL] (d.west) to[out=-160,in=55] (3.north);
		\end{pgfonlayer}
	},
	d5/.pic={
		\begin{pgfonlayer}{nodelayer}
			\node [style=ChFlip] (0) at (-6.25, 1.5) {};
			\node [style=ChFlip] (1) at (-3.75, 1.5) {};
			\node [style=ChFlip] (2) at (-4.35, 0) {};
			\node [style=ChFlip] (3) at (-1, 1.5) {};
			\node [style=none,minimum size=4pt,label={[xshift=-1pt,yshift=-4pt]$a$}] (a) at (-7.85, 2.5) {};
			\node [style=none,minimum size=4pt,label={[xshift=1pt,yshift=-4pt]$b$}] (b) at (-5, 2.5) {};
			\node [style=none,minimum size=4pt,label={[xshift=2pt,yshift=-4pt]$c$}] (c) at (-2.5, 2.5) {};
			\node [style=none,minimum size=4pt,label={[xshift=3pt,yshift=-4pt]$d$}] (d) at (0, 2.5) {};
			\node [style=none,minimum size=4pt,label={[xshift=-1pt,yshift=-4pt]$$}] (ap) at (-7.25, 2.25) {};
			\node [style=none,minimum size=4pt,label={[xshift=1pt,yshift=-4pt]$$}] (bp) at (-4.75, 2.25) {};
			\node [style=none,minimum size=4pt,label={[xshift=2pt,yshift=-4pt]$$}] (cp) at (-2.25, 2.25) {};
		\end{pgfonlayer}
		\begin{pgfonlayer}{edgelayer}
			\draw [style={fermion},PL] (b.south) to [out=15,in=-135] (0.east);
			\draw [style={fermion},PR] (0.west) to [out=-45,in=175] (ap.south) ;
			\draw [style={fermion},PL] (c.south) to[ out=15,in=-135] (1.east) ;
			\draw [style={fermion},PR](1.west) to [out=-45,in=175]  (bp.south) ;
			\draw [style={fermion},PR] (2.east) to[out=15,in=-135] (cp.south);
			\draw [style={fermion},PL] (a.south) to[out=-45,in=175] (2.west)  ;
			\draw [style={fermion},PL] (3.east) to[out=15,in=-135] (d.south);
			\draw [style={fermion},PR] (d.west) to[out=-160,in=55] (3.north);
		\end{pgfonlayer}
	},
	d6/.pic={
		\begin{pgfonlayer}{nodelayer}
			\node [style=ChFlip] (2) at (-4.35, 0) {};
			\node [style=ChFlip] (3) at (-1, 1.5) {};
			\node [style=none,minimum size=4pt,label={[xshift=-1pt,yshift=-4pt]$a$}] (a) at (-7.5, 2.5) {};
			\node [style=none,minimum size=4pt,label={[xshift=1pt,yshift=-4pt]$b$}] (b) at (-5, 2.5) {};
			\node [style=none,minimum size=4pt,label={[xshift=2pt,yshift=-4pt]$c$}] (c) at (-2.5, 2.5) {};
			\node [style=none,minimum size=4pt,label={[xshift=3pt,yshift=-4pt]$d$}] (d) at (0, 2.5) {};
			\node [style=none,minimum size=4pt,label={[xshift=-1pt,yshift=-4pt]$$}] (ap) at (-7.25, 2.25) {};
			\node [style=none,minimum size=4pt,label={[xshift=1pt,yshift=-4pt]$$}] (bp) at (-4.95, 2.25) {};
			\node [style=none,minimum size=4pt,label={[xshift=2pt,yshift=-4pt]$$}] (cp) at (-2.25, 2.25) {};
		\end{pgfonlayer}
		\begin{pgfonlayer}{edgelayer}
			\draw [style={fermion},PR] (b.west) to[out=-135,in=-45] (a.east);
			\draw [style={fermion},PL] (c.west) to[out=-135,in=-45] (b.east);
			\draw [style={fermion},PR] (2.east) to[out=15,in=-135] (c.south);
			\draw [style={fermion},PL] (a.south) to[out=-45,in=175] (2.west) ;
			\draw [style={fermion},PR] (3.east) to[out=15,in=-135] (d.south);
			\draw [style={fermion},PL] (d.west) to[out=-160,in=55] (3.north);
		\end{pgfonlayer}
	},
	d7/.pic={
		\begin{pgfonlayer}{nodelayer}
			\node [style=ChFlip] (2) at (-4.35, 0) {};
			\node [style=ChFlip] (3) at (-1, 1.5) {};
			\node [style=none,minimum size=4pt,label={[xshift=-1pt,yshift=-4pt]$a$}] (a) at (-7.5, 2.5) {};
			\node [style=none,minimum size=4pt,label={[xshift=1pt,yshift=-4pt]$b$}] (b) at (-5, 2.5) {};
			\node [style=none,minimum size=4pt,label={[xshift=2pt,yshift=-4pt]$c$}] (c) at (-2.5, 2.5) {};
			\node [style=none,minimum size=4pt,label={[xshift=3pt,yshift=-4pt]$d$}] (d) at (0, 2.5) {};
			\node [style=none,minimum size=4pt,label={[xshift=-1pt,yshift=-4pt]$$}] (ap) at (-7.25, 2.25) {};
			\node [style=none,minimum size=4pt,label={[xshift=1pt,yshift=-4pt]$$}] (bp) at (-4.95, 2.25) {};
			\node [style=none,minimum size=4pt,label={[xshift=2pt,yshift=-4pt]$$}] (cp) at (-2.25, 2.25) {};
		\end{pgfonlayer}
		\begin{pgfonlayer}{edgelayer}
			\draw [style={fermion},PR] (b.west) to[out=-135,in=-45] (a.east);
			\draw [style={fermion},PL] (c.west) to[out=-135,in=-45] (b.east);
			\draw [style={fermion},PR] (2.east) to[out=15,in=-135] (c.south);
			\draw [style={fermion},PL] (a.south) to[out=-45,in=175] (2.west) ;
			\draw [style={fermion},PL] (3.east) to[out=15,in=-135] (d.south);
			\draw [style={fermion},PR] (d.west) to[out=-160,in=55] (3.north);
		\end{pgfonlayer}
	},
	v-vertex/.pic = {
		\begin{pgfonlayer}{nodelayer}
			\node [style=none,minimum size=4pt,label={[xshift=-2pt,yshift=-4pt]$a$}] (a) at (-7.5, 2.5) {};
			\node [style=none,minimum size=4pt,label={[xshift=2pt,yshift=-4pt]$b$}] (b) at (-4.5, 2.5) {};
			\node [style=none,minimum size=4pt,label={[xshift=-2pt,yshift=-13pt]$d$}] (d) at (-7.5, -0.5) {};
			\node [style=none,minimum size=4pt,label={[xshift=2pt,yshift=-13pt]$c$}] (c) at (-4.5, -0.5) {};
		\end{pgfonlayer}
		\begin{pgfonlayer}{edgelayer}
			\draw [style={fermion},PL, looseness=1.3] (a.south) to[out=-45,in=45] (d.north);
			\draw [style={fermion},PR, looseness=1.3] (b.west) to[out=-135,in=-45] (a.east);
			\draw [style={fermion},PL, looseness=1.3] (c.north) to[out=135,in=-135] (b.south);
			\draw [style={fermion},PR, looseness=1.3] (d.east) to[out=45,in=135] (c.west);
		\end{pgfonlayer}
	},
	u-vertex/.pic = {
		\begin{pgfonlayer}{nodelayer}
			\node [style=none,minimum size=4pt,label={[xshift=-2pt,yshift=-4pt]$a$}] (a) at (-7.5, 2.5) {};
			\node [style=none,minimum size=4pt,label={[xshift=2pt,yshift=-4pt]$b$}] (b) at (-4.5, 2.5) {};
			\node [style=none,minimum size=4pt,label={[xshift=-2pt,yshift=-13pt]$d$}] (d) at (-7.5, -0.5) {};
			\node [style=none,minimum size=4pt,label={[xshift=2pt,yshift=-13pt]$c$}] (c) at (-4.5, -0.5) {};
		\end{pgfonlayer}
		\begin{pgfonlayer}{edgelayer}
			\draw [style={fermion},PL, looseness=1.3] (a.south) to[out=-45,in=-135] (b.south);
			\draw [style={fermion},PR, looseness=1.3] (b.west) to[out=-135,in=-45] (a.east);
			\draw [style={fermion},PL, looseness=1.3] (c.north) to[out=135,in=45] (d.north);
			\draw [style={fermion},PR, looseness=1.3] (d.east) to[out=45,in=135] (c.west);
		\end{pgfonlayer}
	},
	t1/.pic={
		\begin{pgfonlayer}{nodelayer}
			\node [style=ChFlip] (0) at (-6.25, 1.5) {};
			\node [style=ChFlip] (1) at (-6.25, 2.5) {};
			\node [style=ChFlip] (2) at (-3.5, 1.5) {};
			\node [style=ChFlip] (3) at (-1, 1.5) {};
			\node [style=none,minimum size=4pt,label={[xshift=-1pt,yshift=-4pt]$a$}] (a) at (-7.5, 2.5) {};
			\node [style=none,minimum size=4pt,label={[xshift=1pt,yshift=-4pt]$b$}] (b) at (-5, 2.5) {};
			\node [style=none,minimum size=4pt,label={[xshift=-1pt,yshift=-4pt]$$}] (ap) at (-7.25, 2.25) {};
			\node [style=none,minimum size=4pt,label={[xshift=1pt,yshift=-4pt]$$}] (bp) at (-4.75, 2.25) {};
			\node [style=none,minimum size=4pt,label={[xshift=2pt,yshift=-4pt]$c$}] (c) at (-2.5, 2.5) {};
			\node [style=none,minimum size=4pt,label={[xshift=3pt,yshift=-4pt]$d$}] (d) at (0, 2.5) {};
		\end{pgfonlayer}
		\begin{pgfonlayer}{edgelayer}
			\draw [style={fermion},PL] (bp.south)  to[out=15,in=-135] (0.east) ;
			\draw [style={fermion},PR]  (0.west) to[out=-45,in=175]  (ap.south);
			\draw [style={fermion},PR]  (1.east) to[out=15,in=-135] (b.south) ;
			\draw [style={fermion},PL]  (a.south) to[out=-45,in=175]  (1.west);
			\draw [style={fermion},PR] (2.east) to[out=15,in=-135] (c.south);
			\draw [style={fermion},PL] (c.west) to[out=-160,in=85] (2.north);
			\draw [style={fermion},PR] (3.east) to[out=15,in=-135]  (d.south);
			\draw [style={fermion},PL] (d.west) to[out=-160,in=55] (3.north);
		\end{pgfonlayer}
	},
	q1/.pic={
		\begin{pgfonlayer}{nodelayer}
			\node [style=ChFlip] (0) at (-8.5, 1.5) {};
			\node [style=ChFlip] (1) at (-6, 1.5) {};
			\node [style=ChFlip] (2) at (-3.5, 1.5) {};
			\node [style=ChFlip] (3) at (-1, 1.5) {};
			\node [style=none,minimum size=4pt,label={[xshift=-1pt,yshift=-4pt]$a$}] (a) at (-7.5, 2.5) {};
			\node [style=none,minimum size=4pt,label={[xshift=1pt,yshift=-4pt]$b$}] (b) at (-5, 2.5) {};
			\node [style=none,minimum size=4pt,label={[xshift=2pt,yshift=-4pt]$c$}] (c) at (-2.5, 2.5) {};
			\node [style=none,minimum size=4pt,label={[xshift=3pt,yshift=-4pt]$d$}] (d) at (0, 2.5) {};
		\end{pgfonlayer}
		\begin{pgfonlayer}{edgelayer}
			\draw [style={fermion},PR] (0.east) to[out=15,in=-135] (a.south);
			\draw [style={fermion},PL] (a.west) to[out=-160,in=85] (0.north);
			\draw [style={fermion},PR] (1.east) to[out=15,in=-135] (b.south);
			\draw [style={fermion},PL] (b.west) to[out=-160,in=85] (1.north);
			\draw [style={fermion},PR] (2.east) to[out=15,in=-135] (c.south);
			\draw [style={fermion},PL] (c.west) to[out=-160,in=85] (2.north);
			\draw [style={fermion},PR] (3.east) to[out=15,in=-135]  (d.south);
			\draw [style={fermion},PL] (d.west) to[out=-160,in=55] (3.north);
		\end{pgfonlayer}
	},
	O5_ex/.pic={
		\begin{pgfonlayer}{nodelayer}
			\node [style=ChFlip] (0) at (-5, 1) {};
			\node [style=none,minimum size=4pt,label={[xshift=-1pt,yshift=-4pt]$a$}] (a) at (-7.5, 2.5) {};
			\node [style=none,minimum size=4pt,label={[xshift=1pt,yshift=-4pt]$b$}] (b) at (-5, 2.5) {};
			\node [style=none,minimum size=4pt,label={[xshift=2pt,yshift=-4pt]$c$}] (c) at (-2.5, 2.5) {};
		\end{pgfonlayer}
		\begin{pgfonlayer}{edgelayer}
			\draw [style={fermion},PL] (a.south) to[out=-45,in=-135] (b.south);
			\draw [style={fermion},PR] (b.west) to[out=-135,in=-45] (a.east);
			\draw [style={fermion},PR] (0.east) to[out=15,in=-135] (c.south);
			\draw [style={fermion},PL] (c.west) to[out=-160,in=55] (0.north);
		\end{pgfonlayer}
	},
	O6_ex/.pic={
		\begin{pgfonlayer}{nodelayer}
			\node [style=ChFlip] (1) at (-6.25, 1.5) {};
			\node [style=ChFlip] (0) at (-5, 1) {};
			\node [style=none,minimum size=4pt,label={[xshift=-1pt,yshift=-4pt]$a$}] (a) at (-7.5, 2.5) {};
			\node [style=none,minimum size=4pt,label={[xshift=1pt,yshift=-4pt]$b$}] (b) at (-5, 2.5) {};
			\node [style=none,minimum size=4pt,label={[xshift=2pt,yshift=-4pt]$c$}] (c) at (-2.5, 2.5) {};
		\end{pgfonlayer}
		\begin{pgfonlayer}{edgelayer}
			\draw [style={fermion},PL] (b.south) to [out=15,in=-135] (1.east);
			\draw [style={fermion},PR] (1.west) to [out=-45,in=175] (a.south) ;
			\draw [style={fermion},PR] (0.east) to[out=15,in=-135] (c.south);
			\draw [style={fermion},PL] (c.west) to[out=-160,in=55] (0.north);
		\end{pgfonlayer}
	},
		O3_ex/.pic={
		\begin{pgfonlayer}{nodelayer}
			\node [style=ChFlip] (1) at (-6.25, 1.5) {};
			\node [style=ChFlip] (2) at (-6.25, 2.5) {};
			\node [style=ChFlip] (0) at (-5, 1) {};
			\node [style=none,minimum size=4pt,label={[xshift=-1pt,yshift=-4pt]$a$}] (a) at (-7.5, 2.5) {};
			\node [style=none,minimum size=4pt,label={[xshift=1pt,yshift=-4pt]$b$}] (b) at (-5, 2.5) {};
			\node [style=none,minimum size=4pt,label={[xshift=2pt,yshift=-4pt]$c$}] (c) at (-2.5, 2.5) {};
			\node [style=none,minimum size=4pt,label={[xshift=-1pt,yshift=-4pt]$$}] (ap) at (-7.25, 2.25) {};
			\node [style=none,minimum size=4pt,label={[xshift=1pt,yshift=-4pt]$$}] (bp) at (-4.75, 2.25) {};
		\end{pgfonlayer}
		\begin{pgfonlayer}{edgelayer}
			\draw [style={fermion},PL] (bp.south)  to[out=15,in=-135] (1.east) ;
			\draw [style={fermion},PR]  (1.west) to[out=-45,in=175]  (ap.south);
			\draw [style={fermion},PR]  (2.east) to[out=15,in=-135] (b.south) ;
			\draw [style={fermion},PL]  (a.south) to[out=-45,in=175]  (2.west);			%
			\draw [style={fermion},PR] (0.east) to[out=15,in=-135] (c.south);
			\draw [style={fermion},PL] (c.west) to[out=-160,in=55] (0.north);
		\end{pgfonlayer}
	},
	O4_ex/.pic={
		\begin{pgfonlayer}{nodelayer}
			\node [style=ChFlip] (1) at (-6.25, 1.5) {};
			\node [style=ChFlip] (2) at (-6.25, 2.5) {};
			\node [style=ChFlip] (0) at (-5, 1) {};
			\node [style=none,minimum size=4pt,label={[xshift=-1pt,yshift=-4pt]$a$}] (a) at (-7.5, 2.5) {};
			\node [style=none,minimum size=4pt,label={[xshift=1pt,yshift=-4pt]$b$}] (b) at (-5, 2.5) {};
			\node [style=none,minimum size=4pt,label={[xshift=2pt,yshift=-4pt]$c$}] (c) at (-2.5, 2.5) {};
			\node [style=none,minimum size=4pt,label={[xshift=-1pt,yshift=-4pt]$$}] (ap) at (-7.25, 2.25) {};
			\node [style=none,minimum size=4pt,label={[xshift=1pt,yshift=-4pt]$$}] (bp) at (-4.75, 2.25) {};
		\end{pgfonlayer}
		\begin{pgfonlayer}{edgelayer}
			\draw [style={fermion},PL] (bp.south)  to[out=15,in=-135] (1.east) ;
			\draw [style={fermion},PR]  (1.west) to[out=-45,in=175]  (ap.south);
			\draw [style={fermion},PR]  (2.east) to[out=15,in=-135] (b.south) ;
			\draw [style={fermion},PL]  (a.south) to[out=-45,in=175]  (2.west);	
			\draw [style={fermion},PL] (0.east) to[out=15,in=-135] (c.south);
			\draw [style={fermion},PR] (c.west) to[out=-160,in=55] (0.north);
		\end{pgfonlayer}
	},
	O2_ex/.pic={
		\begin{pgfonlayer}{nodelayer}
			\node [style=ChFlip] (0) at (-5, 1) {};
			\node [style=none,minimum size=4pt,label={[xshift=-1pt,yshift=-4pt]$a$}] (a) at (-7.5, 2.5) {};
			\node [style=none,minimum size=4pt,label={[xshift=0pt,yshift=-4pt]$b$}] (b) at (-5, 2.5) {};
			\node [style=none,minimum size=4pt,label={[xshift=2pt,yshift=-4pt]$c$}] (c) at (-2.5, 2.5) {};
			
		\end{pgfonlayer}
		\begin{pgfonlayer}{edgelayer}
			\draw [style={fermion},PR] (b.west) to[out=-135,in=-45] (a.east);
			\draw [style={fermion},PL] (c.west) to[out=-135,in=-45] (b.east);
			\draw [style={fermion},PR] (0.east) to[out=15,in=-135] (c.south);
			\draw [style={fermion},PL] (a.south) to[out=-45,in=175] (0.west) ;
		\end{pgfonlayer}
	},
	O7_ex/.pic={
		\begin{pgfonlayer}{nodelayer}
			\node [style=ChFlip] (0) at (-8.5, 1.5) {};
			\node [style=ChFlip] (1) at (-6, 1.5) {};
			\node [style=ChFlip] (2) at (-3.5, 1.5) {};
			\node [style=none,minimum size=4pt,label={[xshift=-1pt,yshift=-4pt]$a$}] (a) at (-7.5, 2.5) {};
			\node [style=none,minimum size=4pt,label={[xshift=1pt,yshift=-4pt]$b$}] (b) at (-5, 2.5) {};
			\node [style=none,minimum size=4pt,label={[xshift=2pt,yshift=-4pt]$c$}] (c) at (-2.5, 2.5) {};
		\end{pgfonlayer}
		\begin{pgfonlayer}{edgelayer}
			\draw [style={fermion},PR] (0.east) to[out=15,in=-135] (a.south);
			\draw [style={fermion},PL] (a.west) to[out=-160,in=85] (0.north);
			\draw [style={fermion},PR] (1.east) to[out=15,in=-135] (b.south);
			\draw [style={fermion},PL] (b.west) to[out=-160,in=85] (1.north);
			\draw [style={fermion},PR] (2.east) to[out=15,in=-135] (c.south);
			\draw [style={fermion},PL] (c.west) to[out=-160,in=85] (2.north);
		\end{pgfonlayer}
	},
	O8_ex/.pic={
		\begin{pgfonlayer}{nodelayer}
			\node [style=ChFlip] (0) at (-8.5, 1.5) {};
			\node [style=ChFlip] (1) at (-6, 1.5) {};
			\node [style=ChFlip] (2) at (-3.5, 1.5) {};
			\node [style=none,minimum size=4pt,label={[xshift=-1pt,yshift=-4pt]$a$}] (a) at (-7.5, 2.5) {};
			\node [style=none,minimum size=4pt,label={[xshift=1pt,yshift=-4pt]$b$}] (b) at (-5, 2.5) {};
			\node [style=none,minimum size=4pt,label={[xshift=2pt,yshift=-4pt]$c$}] (c) at (-2.5, 2.5) {};
		\end{pgfonlayer}
		\begin{pgfonlayer}{edgelayer}
			\draw [style={fermion},PR] (0.east) to[out=15,in=-135] (a.south);
			\draw [style={fermion},PL] (a.west) to[out=-160,in=85] (0.north);
			\draw [style={fermion},PR] (1.east) to[out=15,in=-135] (b.south);
			\draw [style={fermion},PL] (b.west) to[out=-160,in=85] (1.north);
			\draw [style={fermion},PL] (2.east) to[out=15,in=-135] (c.south);
			\draw [style={fermion},PR] (c.west) to[out=-160,in=85] (2.north);
		\end{pgfonlayer}
	},
	O1_ex/.pic={
		\begin{pgfonlayer}{nodelayer}
			\node [style=ChFlip] (0) at (-6.25, 1.5) {};
			\node [style=ChFlip] (1) at (-3.75, 1.5) {};
			\node [style=ChFlip] (2) at (-5, 0) {};
			\node [style=none,minimum size=4pt,label={[xshift=-1pt,yshift=-4pt]$a$}] (a) at (-7.5, 2.5) {};
			\node [style=none,minimum size=4pt,label={[xshift=-1pt,yshift=-4pt]$$}] (ap) at (-7.25, 2.25) {};
			\node [style=none,minimum size=4pt,label={[xshift=1pt,yshift=-4pt]$b$}] (b) at (-5, 2.5) {};
			\node [style=none,minimum size=4pt,label={[xshift=1pt,yshift=-4pt]$$}] (bp) at (-4.8, 2.4) {};
			\node [style=none,minimum size=4pt,label={[xshift=2pt,yshift=-4pt]$c$}] (c) at (-2.5, 2.5) {};
			\node [style=none,minimum size=4pt,label={[xshift=2pt,yshift=-4pt]$$}] (cp) at (-2.25, 2.25) {};
		\end{pgfonlayer}
		\begin{pgfonlayer}{edgelayer}
			\draw [style={fermion},PL] (b.south) to [out=10,in=-120] (0.east);
			\draw [style={fermion},PR](1.west) to [out=-20,in=-70]  (bp.south) ;
			\draw [style={fermion},PR] (0.west) to [out=-25,in=155] (ap.south) ;
			\draw [style={fermion},PL] (c.south) to[ out=15,in=-135] (1.east) ;
			
			\draw [style={fermion},PR] (2.east) to[out=5,in=-100] (cp.south);
			\draw [style={fermion},PL] (a.south) to[out=-85,in=195] (2.west)  ;
		\end{pgfonlayer}
	}
}

\makeatother
\DeclareMathOperator{\Tr}{Tr}
\def\hc{\ensuremath{\mathrm{h.c.}}}

\definecolor{myred}{HTML}{FF2445}
\definecolor{myred}{HTML}{FF4D67}

\definecolor{myblue}{HTML}{5B7EB7}
\definecolor{myblue}{HTML}{608A91}
\definecolor{myblue}{HTML}{A4A8D1}

\definecolor{mycyan}{HTML}{75A0E6}

\begin{document}
	
	\title{\textbf{\huge On Asymptotic safety in 4D gauge theory with additional dimension=4 operators}}

\author{A.I. {\sc Mukhaeva}}\email{ mukhaeva@theor.jinr.ru}

\affiliation{%
	Joint Institute for Nuclear Research, Joliot-Curie, 6, Dubna 141980, Russia
}

\begin{abstract}
	We study interacting fixed points of simple quantum field theory in four-dimensional $SU(N_c)$  coupled to $N_f$ species of color fermions and $N_f^2$ colorless scalars in the Veneziano limit.
Using  the rich structure of  all possible quartic scalar operators, we find an interacting conformal fixed point with stable vacua and crossovers inbetween. 
We perform calculations in perturbation theory up to four loop in the gauge and three loop in the Yukawa and scalar couplings.  
We also consider anomalous	dimensions for fields, scalar mass squared, and a class of dimension-three operators. 
\end{abstract}
\maketitle

\section{Introduction}
\label{sec:intro}

Studying the asymptotic behavior of dimensionless couplings in quantum field theory (QFT) is crucial for understanding both the Standard Model (SM) and Beyond the Standard Model (BSM) scenarios. Asymptotic freedom \cite{Gross:1973id, Politzer:1973fx}, a key feature of quantum chromodynamics, involves a decrease in coupling value as energy scale increases, leading to an approach toward the Gaussian non-interacting fixed point (FP) in the deep ultraviolet (UV).

Asymptotic safety (AS), an extension of asymptotic freedom, proposed by S. Weinberg \cite{Weinberg:1980gg}, also involves reaching a fixed point in the deep UV, but unlike asymptotic freedom, the fixed point value is nonzero, indicating an interacting theory. Initially introduced as a means to achieve nonperturbative renormalizability for the four-dimensional theory of gravity, AS has been applied to gauge theories to stabilize $U(1)$ gauge couplings at an interacting fixed point, addressing issues such as the Landau pole.

One of the major steps in understanding asymptotic safety was the work of \cite{Litim:2014uca}, where it was shown that some four-dimensional quantum field theories involving $SU(N_c)$  gluons, quarks, and scalars can have weakly coupled ultraviolet fixed points. Furthermore, in Ref.\cite{Bond:2016dvk}, necessary and sufficient conditions for asymptotic safety were given, as well as rigorous no-go theorems analyzing structural renormalization representations of general gauge theories.  It was found that asymptotic safety arises from a loop-level cancellation in which all three types of elementary degrees of freedom - scalars, fermions and gauge fields - are required. The results were also extended to semisimple \cite{Bond:2017lnq}, and supersymmetric gauge theories coupled to matter \cite{Bajc:2016efj,Bond:2017suy,Bajc:2023uls}. The properties of these UV interacting fixed points for matter fields can have implications for low-energy phenomena, leading to various phenomenological predictions. Recent reviews  have considered these implications from a phenomenological standpoint \cite{Eichhorn:2022gku,Bednyakov:2023fmc}.

In this paper we investigate $SU(N_c)$ gauge theory with a weakly interacting fixed point in a model of $N_f$ fermions coupled to $SU(N_c)$ gauge fields and elementary scalars through gauge and Yukawa interactions \cite{Antipin:2013pya,Litim:2014uca}. 

This model also has an explicit global $SU(N_f)_L\times SU(N_f)_R$ flavor symmetry breaking down to  $SU(N_f)$ due to additional operators with $\mathrm{dim}=3$ and $\mathrm{dim}=4$ and is characterized by up to 25 independent couplings. 

In the Veneziano limit with large number of field multiplicities $N_f$ and $N_c$, but constant $N_f/N_c$, the fixed point can be systematically studied in perturbation theory with a small control parameter $\epsilon$, which allows one to extract specific details of the theory. The studies \cite{Litim:2014uca,Bond:2017tbw,Bond:2021tgu,Litim:2015iea}  found fixed points and scaling exponents up to the second order in $\epsilon$, including finite $N_c$ corrections. The papers \cite{Litim:2023tym,Bednyakov:2023asy} provide the fixed point couplings and conformal data up to third order in $\epsilon$ (both for infinite and finite-$N_c$ scenarios) by considering four-loop gauge, three-loop Yukawa, and quartic $\beta$-functions. We provide fixed points and scaling dimensions at 4-3-3 order for gauge-Yukawa-scalar  $\beta$-functions. More generally, we also wish to understand how high-energy fixed points are generated dynamically, what their features are, and whether novel phenomena arise owing to the additional operators breaking the flavour symmetry $SU(N_f)_L\times SU(N_f)_R$ down to diagonal $SU(N_f)$.

The paper is organized as follows. Sections \ref{sec:Model}-\ref{sec:Par-s} provides the main information about the considered model and operators. In Sec.\ref{sec:Comp} we give the details of computation of the renormalization-group (RG) functions. In Sec.\ref{sec:fp} we demonstrate our results for fixed points, anomalous dimensions and scaling exponents.  We conclude in Sec.\ref{sec:res}. The full expressions for beta-functions at $2-1-1$ order can be found in Appendix \ref{sec:appA}. Appendix \ref{sec:appB} discusses the Feynman rules for additional dim=3 and dim=4 operators. 

\section{Model description}
In this section we introduce a simple gauge theory whose interacting fixed points are analyzed  within perturbation theory in the Veneziano limit.

\subsection{Gauge theory}\label{sec:Model}

\begin{table}[htbp]
	\centering
	\begin{tabular}{|c|c|c|c|}
		\hline
		Field  & $SU(N_c)$& $SU_L(N_f)$& $SU_R(N_f)$ \\
		\hline
		$\psi_L$  & $N_c$& $N_f$& $1$ \\
		$\psi_R$  & $N_c$& $1$& $N_f$ \\
		\hline
		$H$ & $1$& $N_f$& $\bar{N_f}$ \\
		\hline
	\end{tabular}
	\caption{Representations of matter fields under gauge $SU(N_c)$ and flavor 
	$SU_L(N_f)$ and $SU_R(N_f)$ groups.}
	\label{tab:LS_groups}
\end{table}

We consider four-dimensional theory with $SU(N_c)$ gauge fields coupled to $N_f$ massless Dirac fermions and a scalar singlet  field $H$ (see Table~\ref{tab:LS_groups}). 
The last is uncharged under the gauge group and carries two flavor indices such that it can be written as an $N_f \times N_f$ complex matrix 
\begin{align}
	H=\sum_{A=0}^{N_f^2-1}(\sigma_A+i\phi_A)T^A,
	\label{eq:H_def}
\end{align}
where $T^A$ are the generalized Gell-Mann matrices, extended to include $T_0 = \frac{1}{\sqrt{2N_f}}\mathbb{I} $  and normalized as $\Tr(T^AT^B ) = \frac{1}{2}\delta_{AB}$ ~\cite{Orlando:2019hte,Antipin:2022naw}.

The intial model \cite{Litim:2014uca} has a global flavor symmetry 
$G = SU_L(N_f) \times SU_R(N_f)$, under which both fermions and scalars transform as 
(see Table~\ref{tab:LS_groups})
\begin{align}
	\psi_L \to L \psi_L, \quad \psi_R \to R \psi_R, \quad H \to LHR^\dagger,  \quad L,R\in SU(N_f)_{L,R}.
\end{align}

The corresponding Lagrangian has the form
\begin{align}
	\mathcal{L}  &= -\frac{1}{4}F^{a\mu\nu}F_{\mu\nu}^a+\mathcal{L}_{gf} + \mathcal{L}_{gh}
	+\Tr(\bar{\psi}i\hat D \psi)+\Tr(\partial^\mu H^\dagger \partial_\mu H )
	- y_1(\Tr[\bar{\psi}_L H \psi_R]+ \hc)	{}\nonumber\\ 
	& \hspace{0.5cm} - m^2\Tr[H^\dagger H] - u\Tr[(H^\dagger H)^2] - v(\Tr[H^\dagger H])^2, 
	\label{eq:Lag}
\end{align}
where $F^a_{\mu\nu}$ is the field strength of the gauge bosons $G^a_\mu$ with $a=1,\ldots,N_c^2-1$. The trace in Eq.~\eqref{eq:Lag} runs over both color and flavor indices.
In what follows we assume a linear $R_\xi$-gauge with $\mathcal{L}_{gf} = - \frac{1}{2\xi} (\partial_\mu G^a_\mu)^2$ together with the corresponding ghost Lagrangian $\mathcal{L}_{gh}$.

In this paper we  consider a class of operators that explicitly breaks the flavor symmetry down to diagonal $SU(N_f)$:
\begin{align}
	\mathcal{L}_{breaking}  &=- y_2(\Tr[\bar{\psi}_L(H^\dagger \psi_R]+ \hc)
	- y_3(\Tr[\bar{\psi}_L  \psi_R]\Tr[H]+\hc)- y_4(\Tr[\bar{\psi}_L  \psi_R]\Tr[H^\dagger]+\hc)\nonumber\\
	& \hspace{0.5cm}
- \vec{\kappa}^{(3)} \cdot \vec{O}^{(3)}
	- \vec{\kappa}^{(4)}_{single} \cdot \vec{O}^{(4)}
	- \vec{\kappa}^{(4)}_{double} \cdot \vec{O}^{(4)}
	- \vec{\kappa}^{(4)}_{triple} \cdot \vec{O}^{(4)} - \vec{\kappa}^{(4)}_{quadruple} \cdot \vec{O}^{(4)}.
	\label{eq:Lag_break}
\end{align}	
One can introduce independent couplings for these dim=3 and dim=4 operators resulting in the following additional contribution to the Lagrangian 
\begin{align}
	\delta \mathcal{L}_3 & = 
	-m_\psi \Tr(\bar\psi \psi)
	- \frac{h_2}{2}
	\left[
	\Tr (H H^\dagger  H) + \hc   
	\right]
	- \frac{h_3}{2}
	\left[
	\Tr (H H^\dagger)\Tr(H) + \hc   
	\right]\nonumber \\
	&- \frac{h_4}{2}
	\left[
	\Tr (H H H)+ \hc   
	\right]
	- \frac{h_5}{2}
	\left[
	\Tr (H H) \Tr(H) + \hc   
	\right]\nonumber \\
	& - \frac{h_6}{2}
	\left[
	\Tr (H H)\Tr(H^\dagger ) + \hc   
	\right]
	- \frac{h_7}{2}
	\left[
	\Tr (H) \Tr(H) \Tr(H) + \hc   
	\right] \nonumber \\
	& 
	- \frac{h_8}{2}
	\left[
	\Tr (H) \Tr(H) \Tr(H^\dagger) + \hc   
	\right]
	= - \vec{\kappa}^{(3)}  \cdot \vec{O}^{(3)}.
	\label{eq:dLag3}
\end{align}

\begin{align}
	\delta \mathcal{L}_4^{s}  &= 
	- s_1
	\left[
	\Tr (H H H H) + \hc   
	\right]
	- s_2
	\left[
	\Tr (H H H^\dagger H^\dagger)    
	\right]
	- s_3
	\left[
	\Tr (H HHH^\dagger) + \hc   
	\right] \nonumber\\ 
	& \hspace{0.5cm}
	= - \vec{\kappa}^{(4)}_{single} \cdot \vec{O}^{(4)}.
	\label{eq:dLag4_sing}
\end{align}
\begin{align}
	\delta \mathcal{L}_4^{d}  &= 
	- d_1
	\left[
	\Tr (H H) \Tr(H^\dagger H^\dagger)  
	\right]
	-d_2
	\left[
	\Tr (H H)\Tr(H H) + \hc   
	\right]\nonumber\\ 
	& \hspace{0.5cm}
	-d_3
	\left[
	\Tr (H H)\Tr(H H^\dagger) + \hc   
	\right]  
	- d_4
	\left[
	\Tr (H H H) \Tr(H)  + \hc   
	\right] \nonumber\\ 
	& \hspace{0.5cm}
	- d_5
	\left[
	\Tr (H H H) \Tr(H^\dagger)  + \hc   
	\right]- d_6
	\left[
	\Tr (H H^\dagger H)  \Tr(H) + \hc   
	\right] \nonumber\\ 
	& \hspace{0.5cm}
	- d_7
	\left[
	\Tr (H H^\dagger H)  \Tr(H^\dagger) + \hc   
	\right] 
	= - \vec{\kappa}^{(4)}_{double} \cdot \vec{O}^{(4)}.
	\label{eq:dLag4_doub}
\end{align}
\begin{align}
	\delta \mathcal{L}_4^t  &= t_1[\Tr(HH)\Tr(H)\Tr(H) + \hc] 
	+ t_2[\Tr(HH)\Tr(H)\Tr(H^\dagger) + \hc] \nonumber\\
	&+t_3[\Tr(HH)\Tr(H^\dagger)\Tr(H^\dagger)+ \hc] 
	+t_4[\Tr(H^\dagger H)\Tr(H)\Tr(H) + \hc]   \nonumber\\
	&+t_5[\Tr(H^\dagger H)\Tr(H)\Tr(H^\dagger) + \hc] = - \vec{\kappa}^{(4)}_{triple} \cdot \vec{O}^{(4)}.
	\label{eq:dLag4_triple}
\end{align}
\begin{align}
	\delta \mathcal{L}_4^q  &= q_1[\Tr(H)\Tr(H)\Tr(H) \Tr(H)+ \hc] + q_2[\Tr(H)\Tr(H)\Tr(H)\Tr(H^\dagger) + \hc] \nonumber\\
	& + q_3[\Tr(H)\Tr(H)\Tr(H^\dagger)\Tr(H^\dagger)+ \hc]= - \vec{\kappa}^{(4)}_{quadruple} \cdot \vec{O}^{(4)}.
	\label{eq:dLag4_quadr}
\end{align}

\subsection{Free parameters and Veneziano limit}\label{sec:Par-s}
As can be observed from Eq.\eqref{eq:Lag} we have four $y_{1,2,3,4}$ Yukawa couplings. 
The couplings mixes under renormalization. One can introduce linear combinations corresponding to the couplings of $\sigma_0(\sigma_a)$, and  $\phi_0(\phi_a)$ to scalar and pseudoscalar flavor (non)-singlet fermion currents, respectively. While the scalar and pseudoscalar couplings do not mix under renormalization, the flavor singlet and flavor non-singlet  couplings do, thus, introducing square roots if expressed in terms of rescaled couplings (see below). Setting $y_{3,4}$ to zero, we make flavor singlet and flavor non-singlet couplings equal. 
Thus, we leave only with two Yukawa couplings. In addition, it should be noted that we redefine the previous couplings and move to new Yukawa defined as $\tilde y_1 \equiv y_1+y_2 $ (corresponds to $\sigma_A$) and  $\tilde y_2 \equiv y_1-y_2 $ (corresponds to $\phi_A$). 

We consider the Veneziano limit \cite{Veneziano:1976wm} with $N_f, N_c \to \infty$. One also introduces a parameter
\begin{align}
	\epsilon & \equiv \frac{N_f}{N_c}-\frac{11}{2}
	\label{eq:eps_def}
\end{align}
that becomes continuous and can take any value between $(-\frac{11}{2},\infty)$.
The benefit of the Veneziano limit is that it allows systematic expansions in a small parameter. 
In our work we suppose that
\begin{align}
	0<\epsilon\leq 1,
\end{align}
and treat it as a small control parameter for perturbativity. 

The model has 23 dimensionless couplings:  gauge coupling $g$, the Yukawa $\tilde y_1$, $\tilde y_2$ and  quartic scalar couplings $u$, $v$, $s_i$, $d_i$, $t_i$, $q_i$. One usually introduces a set of rescaled couplings \cite{tHooft:1973alw}
\begin{align}
	\alpha_g =\frac{g^2 N_c}{(4\pi)^2}, \quad \tilde\alpha_{{ y_{1,2}}} =\frac{{\tilde y_{1,2}}^2 N_c}{(4\pi)^2},\quad
	\alpha_u =\frac{u N_f}{(4\pi)^2}, \quad \alpha_v =\frac{v N_f^2}{(4\pi)^2}.   \label{eq:dim4_couplings_VL}
\end{align}

\begin{align}
	\alpha_{s_i} =\frac{s_i N_f}{(4\pi)^2}, \quad \alpha_{d_i} =\frac{d_i N_f^2}{(4\pi)^2},\quad
	\alpha_{t_i} =\frac{t_i N_f^3}{(4\pi)^2}, \quad \alpha_{q_i} =\frac{q_i N_f^4}{(4\pi)^2}.
	\label{eq:dim4_coupl}
\end{align}
This allows all corrections with positive powers of $N_c$ and $N_f$ appearing in the beta-functions for $\vec{\kappa}$ to be absorbed into the rescaled couplings given in Eqs.~\eqref{eq:dim4_couplings_VL}  and ~\eqref{eq:dim4_coupl}. 

In order to study dimension-3 operators in the Veneziano limit, we rescale the corresponding couplings and the operators as
\begin{align}
	m'_\psi & = m_\psi \sqrt{N_c}, & \alpha_{h_2}  &= h_2 N_f, &
	\alpha_{h_3} &= h_3 N_f^2, & \alpha_{h_4}  &= h_4 N_f, \nonumber\\ \alpha_{h_5}  & = h_5 N_f^2,& 
	\alpha_{h_6}  & = h_6 N_f^2 , & \alpha_{h_7}  &= h_7 N_f^3, &\alpha_{h_8}  &= h_8 N_f^3, \nonumber\\
	O'_1  & = O_1/\sqrt{N_c}, & O'_2 & = O_2/N_f, & O'_3 &= O_3/N_f^2, & O'_4 &= O_4/N_f,\nonumber\\
	O'_5  & = O_5/N_f^2, & O'_6 & = O_6/N_f^2, & O'_7 &= O_7/N_f^3, & O'_8 &= O_8/N_f^3.
	\label{eq:dim3_couplings_VL}
\end{align}

\subsection{Computational technique}\label{sec:Comp}

In perturbation theory the $\beta$-functions can be given as
\begin{align}
	\beta_x \equiv \frac{d \alpha_x}{d \ln \mu} = \beta_x^{(1)} +  \beta_x^{(2)} +  \beta_x^{(3)} + ...,
	\label{eq:beta_expansion}
\end{align}
where $\beta_x^{(n)}$ denotes the $n$-th loop contribution, and $x = \{g,y,u,v\}$.  In Ref.~\cite{Litim:2023tym} they were found for the first time and provided  in the Veneziano limit. Finite-$N_c$ corrections were found in Ref.\cite{Bond:2021tgu}. 
The perturbative RG flow for the gauge-Yukawa-scalar system at NLO$^\prime$  accuracy are
given in Appendix \ref{sec:appA}. Due to the fact that the expressions are very lengthy, we only give this ordering. 
It should be noted that generic $\beta$-functions for the gauge, Yukawa, quartic and trilinear couplings were obtained via  RGBeta software package \cite{Thomsen:2021ncy}. 	

In Appendix \ref{sec:appB} we illustrate Feynman's rules for vertices with real scalars $\phi^a$, in terms of which $H$ has been rewritten as $H = \phi^a T^a$ and $H^\dagger = \phi^a \bar T^a$ with $T^a$ being complex $N_f \times N_f$  matrices ~\cite{Orlando:2019hte,Antipin:2022naw}.

In addition to beta-functions, we also calculated the anomalous dimensions of additional  dim=3 operators as
\begin{align}
	\frac{d}{d \ln \mu} \vec{\kappa} \equiv \dot{\vec{\kappa}}
	= \vec{\beta}_\kappa = \gamma_\kappa(\alpha) \cdot \vec{\kappa}, \qquad \qquad \vec{\kappa} \equiv \vec{\kappa}^{(3)}
\end{align}
and relate the matrix anomalous dimension $\gamma_\kappa$ to the anomalous dimensions of the dimension-3 operators ($d=4 - 2\varepsilon$)
\begin{align}
	\gamma_O(\alpha)  = - 2\varepsilon + \gamma^T_\kappa(\alpha),\qquad
	O \equiv O^{(3)}
	\label{eq:gammaO}
\end{align}
Both $\gamma_O(\alpha)$ and $\gamma_\kappa(\alpha)$ should be finite in the limit $\varepsilon \to 0$. In $d=4$ we have 
\begin{align}
	\gamma_O = \gamma_\kappa^T.
\end{align}
This relation can be applied in two ways.  If we have beta-functions for a closed set of couplings of dimension-1, we can derive the anomalous dimension of the matrix for the corresponding operators.  Conversely, if we know  $\gamma_O$, we can restore the $\overline{\mathrm{MS}}$  beta-functions for operator coupling, e.g., by 
\begin{align}
	\beta_{m_\psi} & = m_\psi (\gamma_O)_{11} + h_i (\gamma_O)_{i1},  \\  
	\beta_{h_i} & = m_\psi (\gamma_O)_{1i} + h_{j} (\gamma_O)_{ij},  \quad i,j=2\ldots 8 
\end{align}
In this paper we use these equations to  compute $\gamma_O$ up to two loops by means of RGBeta.

\section{Interacting fixed points}\label{sec:fp}

In this section, we present our results for fixed points up to the third nontrivial order in the Veneziano limit. These FPs correspond to interacting conformal field theories. In addition, we also find  the universal scaling exponents in their vicinity.

\subsection{Fixed points for marginal operators}

To achieve non-trivial FP, it is necessary to include terms up to two-loop order in the gauge coupling; otherwise, an interacting fixed point cannot be established. Additionally, to investigate the viability of asymptotically safe ultraviolet  fixed points, we must incorporate the Yukawa couplings at least at the leading nontrivial order, which is one loop.
At $2-1-0 $ loop order we have several fixed points. 
\begin{table}[htbp]
	\centering
	\begin{tabular}{|c|c|c|}
		\hline
		$\alpha_g$  & $\tilde\alpha_{y_1}$& $\tilde\alpha_{y_2}$\\
		\hline
		0 & 0& 0 \\
		\hline
		$-\frac{5}{3}$  & 0& $-\frac{4}{3}$  \\
		\hline
		$-\frac{5}{3}$  & $-\frac{4}{3}$ &0  \\
		\hline
		$-\frac{4}{75}$  & 0& 0  \\
		\hline
		$\frac{26}{57}$  & $\frac{4}{19}$& $\frac{4}{19}$  \\
		\hline
	\end{tabular}
	\caption{Fixed points at $2-1-0$ loop order for gauge-Yukawa-scalar}
	\label{tab:fp_210}
\end{table}
Unfortunately, only a single point from this set gives an interacting FP (the last row in Table \ref{tab:fp_210}).  It corresponds to  ``old'' known fixed point  \cite{Litim:2014uca}.

When scalar fields are present, we also need to include the quartic scalar couplings at their leading nontrivial order. This represents the minimal perturbative order required to identify a fully interacting fixed point (NLO$^\prime$).
Further it is not difficult to find leading-order fixed points for the quartic couplings, see Ref.\cite{Litim:2014uca}.

\subsection{Towards additional operators}

These works \cite{Litim:2014uca,Bond:2017tbw,Bond:2021tgu,Litim:2023tym,Bednyakov:2023asy} carry a lot of importance in terms of showing that four-dimensional gauge-Yukawa theories interacting with scalars have UV-fixed points and are asymptotically safe. In our work we would like to investigate the effect of all possible dimension-4 operators on this theory.
With regards to this theory, the questions we would like to answer futher are: 
\begin{itemize}
	\item Do the  dimension-4  scalar couplings change the fixed point structure of this theory?
	\item Can we still attain asymptotic safety with the new dimension-4 operators included?
\end{itemize}

\subsubsection{Fixed points}

The determination of weakly interacting fixed points of $\beta$-function systems (see App.\ref{sec:appA}) does not require much effort. However, taking into account the polynomial character of the beta-function, a large variety of (potentially false) fixed points arises.  
We use the 433-order $\beta$-functions and solve the tower of $\beta$-functions of the additional operator couplings to find the fixed points. 
We determine interacting fixed points up to the complete third order in the small parameter $\epsilon$ and express fixed points as a series expansion
\begin{align}
	\alpha_x^* & = c_{x}^{(1)}\epsilon + c_{x}^{(2)}\epsilon^2 + c_{x}^{(3)}\epsilon^3 + O(\epsilon^4),
	\label{eq:fp_exp}
\end{align}
where $x=\{g,\tilde y_{1,2},u,v,s_i,d_i,t_i,q_i\}$.

The results are accurate up to a given order, which means that corrections in higher loops will generate only subleading terms of order 4. Analytical expressions can be found in Ref \cite{Bednyakov:2023asy}. Here we only give numerical expressions
\begin{align}
	& \alpha_g^* = 0.456 \epsilon + 0.781 \epsilon^2 + 6.610 \epsilon^3 + 24.137 \epsilon^4, \label{eq:ag_fp433}\\
	& \tilde \alpha_{y_1}^* =0.211 \epsilon + 0.508 \epsilon^2 + 3.222 \epsilon^3 + 15.212 \epsilon^4,\label{eq:ay1_fp433}\\
	&  \tilde\alpha_{y_2}^* =0.211 \epsilon + 0.508 \epsilon^2 + 3.222 \epsilon^3 + 15.212 \epsilon^4,\label{eq:ay2_fp433}\\
	&   \alpha_u^* =  0.200 \epsilon + 0.440 \epsilon^2 + 2.693 \epsilon^3+ 12.119 \epsilon^4,\label{eq:au_fp433}\\
	&   \alpha_v^*  = -0.137 \epsilon - 0.632 \epsilon^2 - 4.313 \epsilon^3- 24.147 \epsilon^4,\label{eq:av_fp433}\\
	&\alpha_{ s_i, d_i, t_i, q_i} = 0.
	\label{eq:fp433_quart}
\end{align}

Since we are interested in the Veneziano limit where we have asymptotic safety for four dimensional gauge-Yukawa theories, we have only one fixed point. We note that the additional couplings could have changed the asymptotic safety of the theory, and could have an effect on the gauge-Yukawa interactions. However, what we see in Eq.\eqref{eq:ag_fp433}-\eqref{eq:fp433_quart} is that the theory is still asymptotically safe with the inclusion of dimension-4 operators.

All terms have coefficients of order one.  Also note that all terms of order $\epsilon^3$ appear with the same sign as at order $\epsilon^2$. This means that $\alpha_g$ and $\tilde\alpha_{y_{1,2}} $ remain positive for all $\epsilon$, as they must be and do not impose any restrictions on the domain of applicability.
It is also worth noting that the values for $\tilde\alpha_{y_1}$ and $\tilde\alpha_{y_2}$ coincide. But we should not forget that we have redefined the Yukawa coupling constants. And this corresponds to  $\alpha_{y_1}=(\tilde\alpha_{y_1} +\tilde\alpha_{y_2})/2$ and $\alpha_{y_2}=(\tilde\alpha_{y_1}-\tilde\alpha_{y_2})/2$.

Unfortunately, we were unable to find other interacting fixed points despite the rich range of new operators. 
It is worth noting that analyzing the $\beta$-functions (see App.\ref{sec:appA}), we can identify an independent set of quartic operators $u,v,s_{1,2,3}, d_{1,2,3}$. By independence we mean that the beta-function includes only the listed operators; there is no mixing with the remaining operators $d_{4-7},t_{1-5},q_{1-3}$. This is fulfilled in all-loops.

\subsubsection{Critical exponents}
The critical exponents are universal quantities that determine the behaviour of the flow around the fixed point.
The scaling exponents can be obtained as eigenvalues of the stability matrix
\begin{align}
	M_{ij} = \frac{\partial\beta_i}{\partial\alpha_j}\vert_{\alpha=\alpha^*},
\end{align}
which again we expand as a power series 
\begin{align}
	\theta_j = c_{\theta_j}^{(1)}\epsilon +  c_{\theta_j}^{(2)}\epsilon^2 +c_{\theta_j}^{(3)}\epsilon^3+...
	\label{eq:theta_expansion}
\end{align}
It should be reminded that in a predictive fundamental theory, it is expected that the operators should have a finite number of relevant operators. Also, a negative eigenvalue  has a relevant direction for the associated operator and, similarly, a positive eigenvalue  has an irrelevant direction.

In our case with additional quartic operators, we have the following critical exponents:
\begin{align}
	&   \theta_1 =-\frac{104 \epsilon ^2}{171}+\frac{2296\epsilon^3}{3249}+ \frac{\left(43551640704 \zeta_3+1405590649319-281341851912 \sqrt{23}\right)\epsilon^4}{15643993482},\\
	&   \theta_2=\frac{52\epsilon}{19}+\frac{\left(136601719-22783308 \sqrt{23}\right)\epsilon^2}{4094823}\nonumber\\ 
	& \hspace{0.5cm}+ \frac{5 \left(547695099865475491 - 111718308712462080 \sqrt{23}\right)\epsilon^3}{2692813775855538},\\
	&   \theta_3 =\frac{8}{19}
	\sqrt{2 \left(10+3 \sqrt{23}\right)} \epsilon -\frac{2 \sqrt{20+6 \sqrt{23}} \left(9153184 \sqrt{23}-45155739\right) \epsilon ^2}{16879999}\nonumber\\ 
	& \hspace{0.5cm}+\frac{\left(73205713038142585-15289473238519518 \sqrt{23}\right) \sqrt{20+6 \sqrt{23}} \epsilon ^3}{404906730972633}\nonumber\\ 
	& \hspace{0.5cm}-\frac{1664 \sqrt{20+6 \sqrt{23}} \left(1497 \sqrt{23}-7558\right) \zeta (3) \epsilon ^3}{733913},\\
	&\theta_4= \frac{16 \sqrt{23} \epsilon }{19}+\frac{4 \left(68248487 \sqrt{23}-255832864\right) \epsilon ^2}{31393643}\nonumber\\ 
	& \hspace{0.5cm}\frac{2 \left(10367409687456695575285 \sqrt{23}-48563991480268391192172\right) \epsilon^3}{78238531092995661159},\\
	&   \theta_5 =\frac{8 \epsilon }{19}+\frac{\left(29134-4752 \sqrt{23}\right) \epsilon ^2}{6859}\nonumber\\ 
	& \hspace{0.5cm} +\frac{\epsilon ^3 \left(3875441856 \text{$\zeta $3}-3875441856 \zeta (3)-40698482292 \sqrt{23}+202782332657\right)}{1537657479},\\
	&  \theta_6 =\frac{16 \epsilon }{19}-\frac{2 \left(1065308112 \sqrt{23}-6531289254\right) \epsilon ^2}{1537657479}\nonumber\\ 
	& \hspace{0.5cm}-\frac{2 \left(40698482292 \sqrt{23}-202782332657\right) \epsilon ^3}{1537657479},\\
	&  \theta_7 =\frac{8}{19} \left(1+\sqrt{23}\right) \epsilon+\frac{\left(111882908394 \sqrt{23}-392951182230\right) \epsilon ^2}{35366122017}\nonumber\\ 
	& \hspace{0.5cm}
	+\frac{\left(3356405900191 \sqrt{23}-15267976554869\right) \epsilon ^3}{35366122017},\\
	&    \theta_8 =\frac{8}{19} \left(1+\sqrt{23}\right) \epsilon+\frac{\left(114234118722 \sqrt{23}-421165706166\right) \epsilon ^2}{35366122017}\nonumber\\ 
	& \hspace{0.5cm}
	+\frac{\left(3510431509267 \sqrt{23}-16096577376509\right) \epsilon ^3}{35366122017},\\
	&  \theta_9 =\frac{1}{19} \left(8+8 \sqrt{23}\right)\epsilon+\frac{\left(551514 \sqrt{23}-1959094\right) \epsilon ^2}{157757}\nonumber\\ 
	& \hspace{0.5cm}+\frac{\left(3739339755019 \sqrt{23}-17120803022369\right) \epsilon ^3}{35366122017},\\
	& \theta_{10}=\frac{4}{19} \left(2+\sqrt{20+6 \sqrt{23}}\right) \epsilon \nonumber\\ 
	& \hspace{0.5cm}+ \frac{\left(\left(45155739-9153184 \sqrt{23}\right) \sqrt{20+6 \sqrt{23}}-4922 \left(2376 \sqrt{23}-14567\right)\right) \epsilon ^2}{16879999}\nonumber\\ 
	& \hspace{0.5cm}+\frac{\left(202782332657-40698482292 \sqrt{23}\right) \epsilon ^3}{1537657479}\nonumber\\ 
	& \hspace{0.5cm}+ \frac{\left(73205713038142585-15289473238519518 \sqrt{23}\right) \sqrt{20+6 \sqrt{23}} \epsilon ^3}{809813461945266}\nonumber\\ 
	& \hspace{0.5cm}+\frac{832 \sqrt{\frac{2}{107} \left(931899 \sqrt{23}-4436554\right)} \zeta (3) \epsilon ^3}{6859}
	\label{eq:theta}
\end{align}

For convenience, we provide the numerical results in the Veneziano limit
\begin{align}
	&\theta_1 =-0.608 \epsilon^2 + 0.707\epsilon^3 + 6.947 \epsilon^4,\quad
	\theta_2 =2.737 \epsilon + 6.676 \epsilon^2 + 22.120 \epsilon^3 , \nonumber\\
	&\theta_3 =2.941\epsilon + 1.042 \epsilon^2 + 5.137 \epsilon^3,\quad
	\theta_4 =4.039 \epsilon + 9.107 \epsilon^2 + 38.646 \epsilon^3 ,\nonumber\\
	&\theta_5 =0.421 \epsilon+0.925 \epsilon^2+  4.942 \epsilon^3, \quad
	\theta_6 =0.842 \epsilon + 1.849 \epsilon^2 + 9.884 \epsilon^3,\nonumber\\
	&\theta_7 =2.440 \epsilon + 4.061 \epsilon^2 + 23.434 \epsilon^3,\quad
	\theta_8 =2.440 \epsilon + 3.582 \epsilon^2 + 20.892 \epsilon^3,\nonumber\\
	&\theta_9 =2.440\epsilon + 4.348 \epsilon^2 + 22.972\epsilon^3,\quad
	\theta_{10}= 1.891\epsilon + 1.446\epsilon^2 + 7.511\epsilon^3.
\end{align}

As we can see from Eqs.\eqref{eq:theta}, there is one relevant direction. 

It should be noted that the first $\theta_1$, $\theta_2$, $\theta_3$, $\theta_4$ coincide with our previous calculations \cite{Bednyakov:2023asy}. However, as we have a multitude of additional operators, new critical exponents  $\theta_5$-$\theta_{10}$ appear. Furthermore, it should be noted that we have only 6 new critical exponents, while the model has 21 ($y_2, s_i, d_i, t_i, q_i$) dimensionless coupling constants in addition to the previous 4 ($g, y_1, u, v$). The reason for this is degeneracy. For example, $\theta_6$ has 11-fold degeneracy, while $\theta_8$ has two-fold degeneracy and $\theta_{10}$ has three-fold degeneracy. 	

\subsubsection{Anomalous dimensions}

The results for  anomalous dimensions are provided in our previous work \cite{Bednyakov:2023asy} up to 3-loop order. Here we just specify the results for the scalar  and fermion fields up to two-loops including new couplings 
\begin{align}
	\gamma_H^{(1)}&=\frac{1}{2}\left(\tilde \alpha_{y_1}+\tilde \alpha_{y_2}\right),\\
	\gamma_H^{(2)}&=\frac{1}{16} \left(20 \alpha_g (\tilde\alpha_{y_1}+\tilde\alpha_{y_2})+128 \alpha_{s_1}^2+16
	\alpha_{s_2}^2+32 \alpha_{s_3}^2+32 \alpha_u^2 + 8(\tilde\alpha_{y_1}+\tilde\alpha_{y_2})\right.{}\nonumber\\ 
	& \hspace{0.5cm}\left.-6( \tilde\alpha_{y_1}^2 - \tilde\alpha_{y_2}^2)\epsilon -33( \tilde\alpha_{y_1}^2-\tilde\alpha_{y_2}^2)-12
	\tilde\alpha_{y_1} \tilde\alpha_{y_2} \epsilon -66 \tilde\alpha_{y_1} \tilde\alpha_{y_2}
	\epsilon \right),\\
	\gamma_\psi^{(1)}&=\frac{1}{8} (4 \alpha_g \xi +\left(\tilde\alpha_{y_1}+\tilde\alpha_{y_2}\right) (2 \epsilon +11)),\\
	\gamma_\psi^{(2)}&=\frac{1}{128} \left[-16 \alpha_g^2 (4 \epsilon -\xi  (\xi +8))-32 \alpha_g (2 \epsilon +11) \tilde\alpha_{y_1}+\tilde\alpha_{y_2})\right.{}\nonumber\\ 
	& \hspace{0.5cm}\left.-(2 \epsilon +11) \left(\tilde\alpha_{y_1}^2 (2 \epsilon +35)+2 \tilde\alpha_{y_1} \tilde\alpha_{y_2} (2
	\epsilon +11)+\tilde\alpha_{y_2}^2 (2 \epsilon +35)\right)\right],
\end{align}
where $\xi$ denotes the gauge fixing parameter in the $R_\xi$ gauge. 

Substituting the coefficients into the power series, we have in the Veneziano limit and for $\xi = 0$
\begin{align}
	\gamma_H&=0.2105\epsilon+ 0.4625\epsilon^2,\\
	\gamma_\psi&= 0.5789\epsilon + 0.6243\epsilon^2,
\end{align}
where the results coincide with our previous calculations of \cite{Bednyakov:2023asy} at a fixed point.

We also calculated the anomalous dimensions for dim=3 operators:
\begin{align}
	\gamma_1&=-\frac{4}{19}\epsilon-\left(\frac{14567}{6859}-\frac{2376 \sqrt{23}}{6859}\right)\epsilon^2,\\
	\gamma_2&= \frac{4}{19} \left(1+2 \sqrt{23}\right)\epsilon+ \frac{606162}{6859 \sqrt{23}}-\frac{99745}{6859}\epsilon^2,\\
	\gamma_3&=\frac{4}{19} \left(1+\sqrt{20+6 \sqrt{23}}\right)\epsilon \nonumber\\ 
	& \hspace{0.5cm}+\left( \frac{14567}{6859}
	-\frac{2376 \sqrt{23}}{6859}
	+\frac{\sqrt{2(1475668498887 \sqrt{23}-7061359720318)}}{6859\sqrt{2461}}\right)\epsilon^2,\\
	\gamma_4&=\frac{12}{19} \epsilon +\frac{\left(5543672944446837727632-904219600192605645696 \sqrt{23}\right) }{870095712362665842288}\epsilon ^2.
\end{align}
Numerically we have
\begin{align}
	\gamma_1  & = 
	-0.21053\epsilon  
	-0.46247\epsilon ^2 , \\
	\gamma_2 & = 
	2.22982\epsilon 
	+3.88519\epsilon ^2, \\
	\gamma_3 & = 
	1.68082\epsilon   
	+0.98321\epsilon ^2,\\
	\gamma_4 & = 0.63158\epsilon   
	+1.38742\epsilon ^2.
\end{align}

Here again two comments should be made. First, the anomalous dimensions for $\gamma_1$-$\gamma_3$ coincide with our previous calculations \cite{Bednyakov:2023asy}.  Second, there is a 5-fold degeneracy for $\gamma_4$.

\section{Discussion}\label{sec:res}
Quantum Field Theory \eqref{eq:Lag}-\eqref{eq:Lag_break} we have considered is an important example of an asymptotically safe 4D particle theory with an interacting and perturbatively controlled fixed point at high energies. We extend the previous studies  \cite{Litim:2014uca,Bond:2017tbw,Bond:2021tgu,Litim:2023tym,Bednyakov:2023asy}  by adding additional quartic operators of dimension-3 and dimension-4. 
In this paper, we proposed an attempt to deal with theories with a large number of new operators, where  we investigated the effects of additional dimension-3 and dimension-4 scalar operators on four dimensional gauge-Yukawa theories that have recently been shown to be asymptotically safe in Ref.\cite{Litim:2014uca}. Adding these  interactions could have a devastating effect on the asymptotic safety of the said theory. However, with the hindsight, we can safely say that with the dim=4 interactions that we added to the theory, asymptotic safety is still there with all except one irrelevant operators.

We started by investigating the UV theories up to four loops in perturbation theory, which leads to a third order expansion at the small Veneziano parameter $\epsilon$. The central contributions for this are the four-loop gauge, three-loop Yukawa and quartic $\beta$-functions. So we can confirm that the additional operators do not destroy asymptotic safety.
In addition to the beta-functions, we have computed all fixed points and critical exponents up to third non-trivial order in $\epsilon$.
The results obtained are consistent with unitarity as expected. We also confirm that the additional dim=4  operators are irrelevant. 

The existence of interacting ultraviolet fixed points in particle physics opens many possibilities for model building \cite{Bond:2017wut}. However, for any practical applications it is equally important to understand the size of the corresponding relevant conformal window. Earlier in Ref.\cite{Bednyakov:2023asy}, we investigated the conformal window for the gauge-Yukawa  theory without additional operators. In this work we revisited this question.  Since the fixed point for $\alpha_g, \tilde\alpha_{y_{1,2}}, \alpha_u, \alpha_v$ has not changed, the constraints on the conformal window arising from the couplings perturbativity, vacuum stability and  scaling exponents  have not changed too.

To gain further insight into this theory, the following items might be a topic of further investigation:
\begin{itemize}
	\item Study the prospect for fixed points in the regime where matter field fluctuations dominate over those by the gauge fields (in other words, this is the case when $\epsilon>1$);
	\item Can one derive directly the  beta-function from the functional RG?
\end{itemize}

Overall, we show that the coupling constants of this theory run to a fixed point. All the operators in our theory have irrelevant directions except for the gauge coupling that have the relevant direction. This is a good indicative of a predictive and fundamental theory.

\appendix
\section{Beta-functions at NLO$^\prime $ (2-1-1)}\label{sec:appA}
Following the terminology of \cite{Litim:2014uca} we present the results for $\beta$-functions. 
\begin{align}
	\beta_g&=  \alpha_g^3 \left(\frac{26 \epsilon }{3}+25\right)-\frac{1}{4}\alpha_g^2 (\tilde\alpha_{y_1} +\tilde\alpha_{y_2} )(11+2\epsilon)^2+\frac{4 \epsilon }{3}\alpha_g^2,\\
	\beta_{y_1}&=-6 \alpha_g \tilde\alpha_{y_1}+ \tilde\alpha_{y_1}^2 \left(\epsilon +\frac{15}{2}\right)+\tilde\alpha_{y_1} \tilde\alpha_{y_2} \left(\epsilon +\frac{11}{2}\right),\\
	\beta_{y_2}&=-6 \alpha_g \tilde\alpha_{y_2}+ \tilde\alpha_{y_1} \tilde\alpha_{y_2} \left(\epsilon +\frac{11}{2}\right)+\tilde\alpha_{y_2}^2 \left(\epsilon +\frac{15}{2}\right),
\end{align}

\begin{align}
	\beta_u&=2 \left(\alpha_{s_2}^2+2 \alpha_{s_3}^2\right)+\tilde\alpha_{y_1} \left(\alpha_{s_3}+2 \alpha_u\right)-\frac{3}{4}\tilde\alpha_{y_1} \tilde\alpha_{y_2} \left( 2 \epsilon
	+11\right)+\tilde\alpha_{y_2} (2 \alpha_u-\alpha_{s_3})+8 \alpha_u^2{}\nonumber\\ 
	& \hspace{0.5cm}-\frac{1}{8} (\tilde\alpha_{y_1}^2+\tilde\alpha_{y_2}^2) (2 \epsilon +11),\\
	\beta_{s_1}&=2 \left(4 \alpha_{s_1} \alpha_{s_2}+\alpha_{s_3}^2\right)+\tilde\alpha_{y_1}\left(2 \alpha_{s_1}+\frac{\alpha_{s_3}}{2}\right)+\frac{1}{8}
	\tilde\alpha_{y_1}\tilde\alpha_{y_2} (2 \epsilon +11)+\tilde\alpha_{y_2} \left(2 \alpha_{s_1}-\frac{\alpha_{s_3}}{2}\right){}\nonumber\\ 
	& \hspace{0.5cm}-\frac{1}{16}( \tilde\alpha_{y_1}^2+\tilde\alpha_{y_2}^2) (2 \epsilon
	+11),\\
	\beta_{s_2}&=2 \left(16 \alpha_{s_1}^2+\alpha_{s_2}^2+4 \alpha_{s_3}^2+4 \alpha_{s_2} \alpha_u\right)+2\tilde\alpha_{y_1} \left(\alpha_{s_2}+\alpha_{s_3}\right)+\tilde\alpha_{y_1}\tilde\alpha_{y_2}
	\left(\epsilon +\frac{11}{2}\right){}\nonumber\\ 
	& \hspace{0.5cm}+2 \tilde\alpha_{y_2} (\alpha_{s_2}-\alpha_{s_3})-\frac{1}{4} (\tilde\alpha_{y_1}^2+\tilde\alpha_{y_2}^2) (2 \epsilon +11),\\
	\beta_{s_3}&=\tilde\alpha_{y_1} (2 \alpha_{s_1}+\alpha_{s_2}+2 \alpha_{s_3}+\alpha_u)-\tilde\alpha_{y_2} (2 \alpha_{s_1}+\alpha_{s_2}-2\alpha_{s_3}+\alpha_u){}\nonumber\\ & \hspace{0.5cm}+8 \alpha_{s_3} (2 \alpha_{s_1}
	+\alpha_{s_2}+  \alpha_u)-\frac{1}{4} (\tilde\alpha_{y_1}^2 -\tilde\alpha_{y_2}^2 )(2 \epsilon +11),
\end{align}

\begin{align}
	\beta_v&=4 \left(\alpha_{d_3}^2+4 \alpha_{d_3} \alpha_{s_3}+4 \alpha_{s_1}^2+\alpha_{s_2}^2+3 \alpha_{s_3}^2+2\alpha_u  \alpha_{s_2}+4 \alpha_u\alpha_v+2
	\alpha_{s_2} \alpha_v+3 \alpha_u^2+ \alpha_v^2\right){}\nonumber\\ 
	& \hspace{0.5cm}+\tilde\alpha_{y_1} (2
	\alpha_{d_3}+2 \alpha_v)+\tilde\alpha_{y_2} (2 \alpha_v-2 \alpha_{d_3}),\\
	\beta_{d_1}&=2 \left(2 \alpha_{d_1}^2+4 \alpha_{d_1} \alpha_{s_2}+8 \alpha_{d_2}^2+32 \alpha_{d_2} \alpha_{s_1}+\alpha_{d_3}^2+4 \alpha_{d_3} \alpha_{s_3}+16 \alpha_{s_1}^2+2 \alpha_{s_2}^2+3 \alpha_{s_3}^2\right){}\nonumber\\ 
	& \hspace{0.5cm}+\tilde\alpha_{y_1} (2 \alpha_{d_1}+\alpha_{d_3})+\tilde\alpha_{y_2} (2 \alpha_{d_1}-\alpha_{d_3}),\\
	\beta_{d_2}&=\tilde\alpha_{y_1} (2 \alpha_{d_2}+\frac{\alpha_{d_3}}{2})+\tilde\alpha_{y_2} \left(2
	\alpha_{d_2}-\frac{\alpha_{d_3}}{2}\right){}\nonumber\\ 
	& \hspace{0.5cm}+8 \alpha_{d_2} \alpha_{s_2}+\alpha_{d_3}^2+4 \alpha_{d_3} \alpha_{s_3}+8 \alpha_{s_1} \alpha_{s_2}+8 \alpha_{s_1} \alpha_u+3 \alpha_{s_3}^2 + 8 \alpha_{d_1} \alpha_{d_2}+16\alpha_{d_1}  \alpha_{s_1}),\\
	\beta_{d_3}&=4 (\alpha_{d_1} +2 \alpha_{d_2})(\alpha_{d_3}+2 \alpha_{s_3})+(2 \alpha_{d_3}+3 \alpha_{s_3}) (2
	\alpha_{s_1}+\alpha_{s_2})){}\nonumber\\ 
	& \hspace{0.5cm}+\tilde\alpha_{y_1} (\alpha_{d_1}+2 \alpha_{d_2}+2\alpha_{d_3}+\alpha_v)-\tilde\alpha_{y_2}
	(\alpha_{d_1}+2 \alpha_{d_2}-2 \alpha_{d_3}+\alpha_v){}\nonumber\\ 
	& \hspace{0.5cm}+\alpha_u (8 \alpha_{d_3}+12 \alpha_{s_3})+4 \alpha_v
	(\alpha_{d_3}+2 \alpha_{s_3}),\\
	\beta_{d_4}&=\frac{1}{2} \tilde\alpha_{y_1} (4 \alpha_{d_4}+\alpha_{d_5}+\alpha_{d_6})+\frac{1}{2} \tilde\alpha_{y_2} (4 \alpha_{d_4}-\alpha_{d_5}-\alpha_{d_6})+6 \alpha_{d_4} \alpha_{s_2}+4 \alpha_{s_3} (\alpha_{d_6}+\alpha_{s_3}){}\nonumber\\ 
	& \hspace{0.5cm}+8 \alpha_{d_7} \alpha_{s_1}+8
	\alpha_{s_1} \alpha_{s_2}+16 \alpha_{s_1} \alpha_u,\\
	\beta_{d_5}&=\frac{1}{2} \tilde\alpha_{y_1} (\alpha_{d_4}+4 \alpha_{d_5}+\alpha_{d_7})-\frac{1}{2} \tilde\alpha_{y_2} (\alpha_{d_4}-4 \alpha_{d_5}+\alpha_{d_7})+6 \alpha_{d_5} \alpha_{s_2}+8 \alpha_{d_6} \alpha_{s_1}{}\nonumber\\ 
	& \hspace{0.5cm}+4 \alpha_{d_7} \alpha_{s_3}+8 \alpha_{s_1}
	\alpha_{s_3}+6 \alpha_{s_2} \alpha_{s_3},\\
	\beta_{d_6}&=\frac{1}{2} \tilde\alpha_{y_1} (3 \alpha_{d_4}+4 \alpha_{d_6}+3 \alpha_{d_7})-\frac{1}{2} \tilde\alpha_{y_2} (3 \alpha_{d_4}-4 \alpha_{d_6}+3
	\alpha_{d_7}){}\nonumber\\ 
	& \hspace{0.5cm}+2 \alpha_{s_3} (6 \alpha_{d_4}+4 \alpha_{d_7}+12 \alpha_{s_1}+5 \alpha_{s_2})+24 \alpha_{d_5} \alpha_{s_1}+6
	\alpha_{d_6} \alpha_{s_2}+8 \alpha_u (\alpha_{d_6}+2 \alpha_{s_3}),\\
	\beta_{d_7}&=2 \left(12 \alpha_{d_4} \alpha_{s_1}+6 \alpha_{d_5} \alpha_{s_3}+4 \alpha_{d_6} \alpha_{s_3}+3 \alpha_{d_7} \alpha_{s_2}+16
	\alpha_{s_1}^2+2 \alpha_{s_2}^2+6 \alpha_{s_3}^2\right){}\nonumber\\ 
	& \hspace{0.5cm}+\frac{1}{2} \tilde\alpha_{y_1} (3 \alpha_{d_5}+3 \alpha_{d_6}+4 \alpha_{d_7})-\frac{1}{2} \tilde\alpha_{y_2} (3 \alpha_{d_5}+3 \alpha_{d_6}-4 \alpha_{d_7})+\alpha_u (8 \alpha_{d_7}+12 \alpha_{s_2}),
\end{align}

\begin{align}
	\beta_{t_1}&=2 \left[2 \alpha_{d_1} (3 \alpha_{d_4}+2 \alpha_{s_1}+\alpha_{t_1})+4 \alpha_{d_2} \alpha_{d_7}+4 \alpha_{d_2} \alpha_{t_3}+\alpha_{d_3} (2 \alpha_{d_6}+\alpha_{s_3}+\alpha_{t_4})\right.{}\nonumber\\ 
	& \hspace{0.5cm}\left.+3 \alpha_{d_4} \alpha_{d_7}+3 \alpha_{d_4} \alpha_{s_2}+\alpha_{d_6}^2+4 \alpha_{d_6} \alpha_{s_3}+8 \alpha_{d_7} \alpha_{s_1}+8 \alpha_{s_1} \alpha_{t_3}+2 \alpha_{s_2}
	\alpha_{t_1}+2 \alpha_{s_3} \alpha_{t_4}\right]{}\nonumber\\ 
	& \hspace{0.5cm}+\alpha_u (8 \alpha_{d_2}+12 \alpha_{d_4})+\frac{1}{2} \tilde\alpha_{y_1} (4
	\alpha_{t_1}+\alpha_{t_2}+\alpha_{t_4})+\frac{1}{2} \tilde\alpha_{y_2} (4 \alpha_{t_1}-\alpha_{t_2}-\alpha_{t_4}),\\
	\beta_{t_2}&=2 (2 \alpha_{d_1} +4 \alpha_{d_2} )(3 \alpha_{d_5}+\alpha_{d_6}+\alpha_{s_3}+\alpha_{t_2})+4 \alpha_{d_3} \alpha_{d_7}{}\nonumber\\ 
	& \hspace{0.5cm}+2 \alpha_{d_3} \alpha_{s_2}+2 \alpha_{d_3} \alpha_{t_5}+3
	\alpha_{d_4} \alpha_{d_6}+3 \alpha_{d_4} \alpha_{s_3}+3 \alpha_{d_5} \alpha_{d_7}+12 \alpha_{d_5} \alpha_{s_1}+3
	\alpha_{d_5} \alpha_{s_2}{}\nonumber\\ 
	& \hspace{0.5cm}+2 \alpha_{d_6} \alpha_{d_7}+8 \alpha_{d_6} \alpha_{s_1}+3 \alpha_{d_6} \alpha_{s_2}+7
	\alpha_{d_7} \alpha_{s_3}+8 \alpha_{s_1} \alpha_{t_2}+2 \alpha_{s_2} \alpha_{t_2}+4 \alpha_{s_3} \alpha_{t_5}){}\nonumber\\ 
	& \hspace{0.5cm}+12
	\alpha_{d_5} \alpha_u+\tilde\alpha_{y_1} (\alpha_{t_1}+2 \alpha_{t_2}+\alpha_{t_3}+\alpha_{t_5})-\tilde\alpha_{y_2} (\alpha_{t_1}-2 \alpha_{t_2}+\alpha_{t_3}+\alpha_{t_5}),\\
	\beta_{t_3}&=2 \left[2 \alpha_{d_1} \alpha_{d_7}+2 \alpha_{d_1} \alpha_{t_3}+4 \alpha_{d_2} (3 \alpha_{d_4}+2 \alpha_{s_1}+\alpha_{t_1})+\alpha_{d_3} (2 \alpha_{d_6}+\alpha_{s_3}+\alpha_{t_4})\right.{}\nonumber\\ 
	& \hspace{0.5cm}\left.+12 \alpha_{d_4} \alpha_{s_1}+3 \alpha_{d_5} \alpha_{d_6}+3
	\alpha_{d_5} \alpha_{s_3}+3 \alpha_{d_6} \alpha_{s_3}+\alpha_{d_7}^2+3 \alpha_{d_7} \alpha_{s_2}+8 \alpha_{s_1}
	\alpha_{t_1}{}\right.\nonumber\\ 
	& \hspace{0.5cm}\left.+2 \alpha_{s_2} \alpha_{t_3}+2 \alpha_{s_3} \alpha_{t_4}\right]+4 \alpha_{d_1} \alpha_u+\frac{1}{2} \tilde\alpha_{y_1}(\alpha_{t_2}+4 \alpha_{t_3}+\alpha_{t_4})-\frac{1}{2} \tilde\alpha_{y_2} (\alpha_{t_2}-4 \alpha_{t_3}+\alpha_{t_4}),\\
	\beta_{t_4}&=2 [2 \alpha_{d_3} (3 \alpha_{d_4}+\alpha_{d_7}+2 \alpha_{s_1}+\alpha_{t_1}+\alpha_{t_3})+9 \alpha_{d_4} (\alpha_{d_5}+\alpha_{s_3})+12 \alpha_{d_5} \alpha_{s_1}+3 \alpha_{d_6} \alpha_{d_7}{}\nonumber\\ 
	& \hspace{0.5cm}+4 \alpha_{d_6} \alpha_{s_2}+5 \alpha_{d_7}
	\alpha_{s_3}+2 \alpha_{s_2} \alpha_{t_4}+4 \alpha_{s_3} \alpha_{t_1}+4 \alpha_{s_3} \alpha_{t_3}]+4 \alpha_u
	(\alpha_{d_3}+3 \alpha_{d_6}+2 \alpha_{t_4}){}\nonumber\\ 
	& \hspace{0.5cm}+4 \alpha_v (2 \alpha_{d_6}+\alpha_{s_3}+\alpha_{t_4})+\tilde\alpha_{y_1}
	(\alpha_{t_1}+\alpha_{t_3}+2 \alpha_{t_4}+\alpha_{t_5})-\tilde\alpha_{y_2} (\alpha_{t_1}+\alpha_{t_3}-2 \alpha_{t_4}+\alpha_{t_5}),\\
	\beta_{t_5}&=4 \alpha_{d_3} (3 \alpha_{d_5}+\alpha_{d_6}+\alpha_{s_3}+\alpha_{t_2})+9 \alpha_{d_4}^2+24 \alpha_{d_4} \alpha_{s_1}+9
	\alpha_{d_5}^2+18 \alpha_{d_5} \alpha_{s_3}+3 \alpha_{d_6}^2{}\nonumber\\ 
	& \hspace{0.5cm}+10 \alpha_{d_6} \alpha_{s_3}+3 \alpha_{d_7}^2+4 \alpha_v (2
	\alpha_{d_7}+\alpha_{s_2}+\alpha_{t_5})+8 \alpha_{d_7} \alpha_{s_2}+\alpha_u (12 \alpha_{d_7}+8 \alpha_{t_5}){}\nonumber\\ 
	& \hspace{0.5cm}+4
	\alpha_{s_2} \alpha_{t_5}+8 \alpha_{s_3} \alpha_{t_2}+\tilde\alpha_{y_1} (\alpha_{t_2}+\alpha_{t_4}+2 \alpha_{t_5})-\tilde\alpha_{y_2} (\alpha_{t_2}+\alpha_{t_4}-2 \alpha_{t_5}),
\end{align}

\begin{align}
	\beta_{q_1}&=6 \alpha_{d_4} (\alpha_{d_7}+2 \alpha_{t_3})+2 \alpha_{d_6}^2+4 \alpha_{d_6} \alpha_{t_4}+4 \alpha_{d_7}\alpha_{t_1}+\frac{1}{2} \tilde\alpha_{y_1} (4 \alpha_{q_1}+\alpha_{q_2}){}\nonumber\\ 
	& \hspace{0.5cm}+\frac{1}{2} \tilde\alpha_{y_2} \left(4 \alpha_{q_1}-\alpha_{q_2}\right)+8
	\alpha_{s_1} \alpha_{t_3}+2 \alpha_{s_3} \alpha_{t_4}+4 \alpha_{t_1} \alpha_{t_3}+4 \alpha_{t_1} \alpha_u+\alpha_{t_4}^2,\\
	\beta_{q_2}&=6 \alpha_{d_4} (3 \alpha_{d_5}+\alpha_{d_6}+2 \alpha_{t_2})+6 \alpha_{d_5} (\alpha_{d_7}+2 (\alpha_{t_1}+\alpha_{t_3}))+10
	\alpha_{d_6} \alpha_{d_7}+4 \alpha_{d_6} \alpha_{t_1}{}\nonumber\\ 
	& \hspace{0.5cm}+4 \alpha_{d_6} \alpha_{t_3}+8 \alpha_{d_6} \alpha_{t_5}+4
	\alpha_{d_7} \alpha_{t_2}+8 \alpha_{d_7} \alpha_{t_4}+\tilde\alpha_{y_1} (2 \alpha_{q_1}+2 \alpha_{q_2}+\alpha_{q_3}){}\nonumber\\ 
	& \hspace{0.5cm}-\tilde\alpha_{y_2} (2 \alpha_{q_1}-2 \alpha_{q_2}+\alpha_{q_3})+8 \alpha_{s_1} \alpha_{t_2}+4 \alpha_{s_2} \alpha_{t_4}+4
	\alpha_{s_3} \alpha_{t_1}+4 \alpha_{s_3} \alpha_{t_3}{}\nonumber\\ 
	& \hspace{0.5cm}+4 \alpha_{s_3} \alpha_{t_5}+4 \alpha_{t_1} \alpha_{t_2}+4
	\alpha_{t_2} \alpha_{t_3}+4 \alpha_{t_2} \alpha_u+4 \alpha_{t_4} \alpha_{t_5},\\
	\beta_{q_3}&=2 \left[9 \alpha_{d_4}^2+12 \alpha_{d_4} \alpha_{t_1}+9 \alpha_{d_5}^2+6 \alpha_{d_5} (\alpha_{d_6}+2 \alpha_{t_2})+3 \alpha_{d_6}^2+4 \alpha_{d_6} \alpha_{t_2}+4 \alpha_{d_6} \alpha_{t_4}\right.{}\nonumber\\ 
	& \hspace{0.5cm}\left.+5 \alpha_{d_7}^2+4 \alpha_{d_7} \alpha_{t_3}+8 \alpha_{d_7}
	\alpha_{t_5}+8 \alpha_{s_1} \alpha_{t_1}+4 \alpha_{s_2} \alpha_{t_5}+4 \alpha_{s_3} \alpha_{t_2}+2 \alpha_{s_3}
	\alpha_{t_4}+2 \alpha_{t_1}^2\right.{}\nonumber\\ 
	& \hspace{0.5cm}\left.+2 \alpha_{t_2}^2+2 \alpha_{t_3}^2+\alpha_{t_4}^2+2 \alpha_{t_5}^2\right]+\tilde\alpha_{y_1} (3\alpha_{q_2}+2 \alpha_{q_3})-\tilde\alpha_{y_2} (3 \alpha_{q_2}+2 \alpha_{q_3})+8 \alpha_{t_3} \alpha_u.
\end{align}

Expressions for 1-loop dim=3 beta-functions have the form
\begin{align}
	\beta_{m'_\psi}&=m'_\psi \left(\frac{1}{4} (2 \epsilon +11) (3\tilde\alpha_{y_1}-\tilde\alpha_{y_2})-3 \alpha_g\right),\\
	\beta_ {\alpha_ {h_ 2}} & = \alpha_ {h_ 2}\left (6\alpha_ {s_ 2} + 8\alpha_ {s_ 3} + 
	8\alpha_u + \frac {5\tilde\alpha_ {y_ 1}} {2} + \frac {\tilde\alpha_ {y_ 2}} {2} \right) + \frac {3} {2}
	\alpha_ {h_ 4} (16\alpha_ {s_ 1} + 
	8\alpha_ {s_ 3} + \tilde\alpha_ {y_ 1} - \tilde\alpha_ {y_ 2}) {}\nonumber\\ 
	& \hspace{0.5cm}- 
	m' _\psi\sqrt {\tilde\alpha_ {y_ 1}} (2\epsilon + 
	11) (3\tilde\alpha_ {y_ 1} + \tilde\alpha_ {y_ 2}), \\
	\beta_ {\alpha_ {h_ 3}} & = 
	6\alpha_ {h_ 4} (2\alpha_ {d_ 3} + 3\alpha_ {d_ 4} + 
	3\alpha_ {d_ 5} + 4\alpha_ {s_ 1} + 3\alpha_ {s_ 3}) {}\nonumber\\ 
	& \hspace{0.5cm}+ 
	2\alpha_ {h_ 2} (2\alpha_ {d_ 3} + 3\alpha_ {d_ 6} + 
	3\alpha_ {d_ 7} + 4\alpha_ {s_ 2} + 5\alpha_ {s_ 3} + 
	6\alpha_u + 4\alpha_v) \nonumber\\ 
	& \hspace{0.5cm}+ \alpha_ {h_ 5} (4\alpha_ {d_ 3} + 
	8\alpha_ {s_ 3} + \tilde\alpha_ {y_ 1} - \tilde\alpha_ {y_ 2}){}\nonumber\\ 
	& \hspace{0.5cm} + \alpha_ {h_ 6} (4\alpha_ {d_ 3} + 
	8\alpha_ {s_ 3} + \tilde\alpha_ {y_ 1} - \tilde\alpha_ {y_ 2}) + \alpha_ {h_ 3} (4\alpha_ {s_ 2} + 
	8\alpha_u + 4\alpha_v + 
	2\tilde\alpha_ {y_ 1} + \tilde\alpha_ {y_ 2}),  \\
	\beta_ {\alpha_ {h_ 4}} & = \frac {1} {2}\alpha_ {h_ 2} (16\alpha_ {s_ 1} + 
	8\alpha_ {s_ 3} + \tilde\alpha_ {y_ 1} - \tilde\alpha_ {y_ 2}) + \frac {3} {2}\alpha_ {h_ 4} (4\alpha_{s_ 2} + \tilde\alpha_ {y_ 1} + \tilde\alpha_ {y_ 2}) {}\nonumber\\ 
	& \hspace{0.5cm}- 
	m' _\psi\sqrt {\tilde\alpha_ {y_ 1}} (2\epsilon + 
	11) (\tilde\alpha_ {y_ 1} - \tilde\alpha_ {y_ 2}), \\
	\beta_ {\alpha_ {h_ 5}} & = 
	6\alpha_ {h_ 4} (2\alpha_ {d_ 1} + \alpha_ {d_ 7} + \alpha_ {s_ 2} + 
	2\alpha_u) + \alpha_ {h_ 5}\left(4\alpha_ {d_ 1} + 
	4\alpha_ {s_ 2} + \frac {3 (\tilde\alpha_ {y_ 1} + \tilde\alpha_ {y_ 2})} {2} \right)\nonumber\\ 
	& \hspace{0.5cm} + 
	2\alpha_ {h_ 2} (4\alpha_ {d_ 2} + 2\alpha_ {d_ 3} + 
	3\alpha_ {d_ 4} + 2\alpha_ {d_ 6} + 8\alpha_ {s_ 1} + 
	4\alpha_ {s_ 3}) + \frac {1} {2}\alpha_ {h_ 6} (16\alpha_ {d_ 2} + 32\alpha_ {s_ 1} + \tilde\alpha_ {y_ 1} - \tilde\alpha_ {y_ 2})\nonumber\\ 
	& \hspace{0.5cm} + \frac {1} {2}\alpha_ {h_ 3} (4\alpha_ {d_ 3} + 
	8\alpha_ {s_ 3} + \tilde\alpha_ {y_ 1} - \tilde\alpha_ {y_ 2}), \\
	\beta_ {\alpha_ {h_ 6}} & = \alpha_ {h_ 2} (4\alpha_ {d_ 1} + 
	4\alpha_ {d_ 3} + 6\alpha_ {d_ 5} + 4\alpha_ {d_ 7} + 
	6\alpha_ {s_ 2} + 6\alpha_ {s_ 3}) + \alpha_ {h_ 6}
	\left (4\alpha_ {d_ 1} + 
	4\alpha_ {s_ 2} + \frac {3 (\tilde\alpha_ {y_ 1} + \tilde\alpha_ {y_ 2})} {2} \right) \nonumber\\ 
	& \hspace{0.5cm}+ 
	6\alpha_ {h_ 4} (4\alpha_ {d_ 2} + \alpha_ {d_ 6} + 
	4\alpha_ {s_ 1} + \alpha_ {s_ 3}) {}\nonumber\\ 
	& \hspace{0.5cm}+ \frac {1} {2}\alpha_ {h_ 5} (16\alpha_ {d_ 2} + 
	32\alpha_ {s_ 1} + \tilde\alpha_ {y_ 1} - \tilde\alpha_ {y_ 2}) + \frac {1} {2}\alpha_ {h_ 3} (4\alpha_ {d_3} + 8\alpha_ {s_ 3} + \tilde\alpha_ {y_ 1} - \tilde\alpha_ {y_ 2}),  \\
	\beta_ {\alpha_ {h_ 7}} & = 
	6\alpha_ {d_ 4}\alpha_ {h_ 2} + 
	4\alpha_ {h_ 6} (3\alpha_ {d_ 4} + 
	2\alpha_ {s_ 1} + \alpha_ {t_ 1}) + 
	4\alpha_ {h_ 2} (\alpha_ {d_ 6} + \alpha_ {t_ 1} + \alpha_ {t_ 4}) + 2\alpha_ {h_ 3} (2\alpha_ {d_ 6} + \alpha_ {s_ 3} + \alpha_ {t_ 4}) \nonumber\\ 
	& \hspace{0.5cm}+ 6\alpha_ {h_ 4} (\alpha_ {d_ 7} + 2\alpha_ {t_ 3}) + 
	4\alpha_ {h_ 5} (\alpha_ {d_ 7} + \alpha_ {t_ 3} + \alpha_u) + \frac {3} {2}\alpha_ {h_ 7} (\tilde\alpha_ {y_ 1} + \tilde\alpha_ {y_ 2}) + \frac {1} {2}\alpha_ {h_ 8} (\tilde\alpha_ {y_ 1} - \tilde\alpha_ {y_ 2}),  \\
	\beta_ {\alpha_ {h_ 8}} & = 
	6\alpha_ {h_4} (3\alpha_ {d_4} + 
	3\alpha_ {d_5} + \alpha_ {d_6} + 2\alpha_ {t_1} + 
	2\alpha_ {t_2}) + 
	4\alpha_ {h_5} (3\alpha_ {d_4} + 
	3\alpha_ {d_5} + \alpha_ {d_6} + 
	2\alpha_ {s_1} + \alpha_ {s_3} + \alpha_ {t_ 1} + \alpha_ {t_2}) \nonumber\\ 
	& \hspace{0.5cm}+ 2\alpha_ {h_2} (3\alpha_ {d_5} + 3\alpha_ {d_6} + 
	5\alpha_ {d_7} + 2\alpha_ {t_2} + 2\alpha_ {t_ 3} + 
	2\alpha_ {t_4} + 4\alpha_ {t_5}) {}\nonumber\\ 
	& \hspace{0.5cm}+ 
	4\alpha_ {h_6} (3\alpha_ {d_5} + \alpha_ {d_ 6} + \alpha_{d_7} 
	+ \alpha_ {s_3} + \alpha_ {t_2} + \alpha_ {t_3} + \alpha_u) \nonumber\\ 
	& \hspace{0.5cm}+ 
	2\alpha_ {h_3} (2\alpha_ {d_6} + 4\alpha_ {d_7} + 
	2\alpha_ {s_2} + \alpha_ {s_3} + \alpha_ {t_4} + 
	2\alpha_ {t_5}) + \frac {3} {2}\alpha_ {h_7} (\tilde\alpha_ {y_1} - \tilde\alpha_ {y_2}) + \frac {1} {2}\alpha_ {h_8} (5\tilde\alpha_ {y_1} + \tilde\alpha_ {y_2}).
\end{align}

\section{Feynman rules for additional operators}\label{sec:appB}
In order to show the flavor flows, we rewrite the Lagrangian \eqref{eq:Lag} in terms of real scalars $\phi^a$ using a decomposition ($a=1,\ldots,2 N_f^2$) \cite{Bednyakov:2021ojn}
\begin{align}
	H = \phi^a T^a, \qquad H^\dagger = \phi^a \bar T^a,
	\qquad \bar T^a \equiv T^{a\dagger}
	\label{eq:H_decomposition}
\end{align}
with $T^a$ being complex $N_f \times N_f$ matrices
normalized as \begin{align}
	\Tr(T^a \bar T^b) + \Tr(T^b \bar T^a)  = \delta^{ab}
\end{align}
and satisfying the following identities
\begin{align}
	T^a_{ij} T^a_{kl} = \bar T^a_{ij} \bar T^a_{kl} = 0, \qquad 
	T^a_{ij} \bar T^a_{kl} = \delta_{il} \delta_{jk}.
	\label{eq:Tmatrix_rules}
\end{align}
Such a decomposition gives rise to the Feynman rules for the vertices involving $\phi^a$.
Feynman rules for dimension-three operators are illustrated in Fig.\ref{fig:O3_feynman_rules}-\ref{fig:O3_feynman_rules_2}.

\begin{figure*}[htbp]
	\begin{align*}
		\begin{tikzpicture}
			\begin{pgfonlayer}{nodelayer}
				\node [style=Op] (0) at (-6, 1) {};
				\node [style=none,label={[xshift=5pt,yshift=-1pt]$O_1$}] (Op) at (-7, 1) {};
				\node [style=none,label={[xshift=+10pt,yshift=-2pt]$\psi_{\alpha}^{i}$}] (1) at (-4.5, 2.5) {};
				\node [style=none,label={[xshift=+10pt,yshift=-14pt]$\psi_{\beta}^{j}$}] (2) at (-4.5, -0.5) {};
			\end{pgfonlayer}
			\begin{pgfonlayer}{edgelayer}
				\draw [style=fermion] (2.center) to (0);
				\draw [style=fermion] (0) to (1.center);
			\end{pgfonlayer}
		\end{tikzpicture}
		& = -i \delta_{\alpha\beta} \delta_{ij} (P_L + P_R) 
		\hspace{1cm} \Rightarrow \hspace{1cm} 
		\begin{tikzpicture}
			\begin{pgfonlayer}{nodelayer}
				\node [style=Op] (0) at (-6, 1) {};
				\node [style=none,label={[xshift=10pt,yshift=-2pt]$\psi_{\alpha,L}^{i}$}] (1) at (-4.5, 2.5) {};
				\node [style=none,label={[xshift=10pt,yshift=-14pt]$\psi_{\beta,R}^{j}$}] (2) at (-4.5, -0.5) {};
			\end{pgfonlayer}
			\begin{pgfonlayer}{edgelayer}
				\draw [style=fermion,PR] (2.center) to (0);
				\draw [style=fermion,PL] (0) to (1.center);
			\end{pgfonlayer}
		\end{tikzpicture}
		+
		\begin{tikzpicture}
			\begin{pgfonlayer}{nodelayer}
				\node[style=none] (x) at (-8,1) {};
				\node [style=Op] (0) at (-6, 1) {};
				\node [style=none,label={[xshift=10pt,yshift=-2pt]$\psi_{\alpha,R}^{i}$}] (1) at (-4.5, 2.5) {};
				\node [style=none,label={[xshift=10pt,yshift=-14pt]$\psi_{\beta,L}^{j}$}] (2) at (-4.5, -0.5) {};
			\end{pgfonlayer}
			\begin{pgfonlayer}{edgelayer}
				\draw [style=fermion,PL] (2.center) to (0);
				\draw [style=fermion,PR] (0) to (1.center);
			\end{pgfonlayer}
		\end{tikzpicture},
		\\
		\begin{tikzpicture}
			\begin{pgfonlayer}{nodelayer}
				\node [style=Op] (0) at (-6, 1) {};
				\node [style=none,label={[xshift=5pt,yshift=-1pt]$O_2$}] (3) at (-7, 1) {};
				\node [style=none,label={[xshift=10pt,yshift=0pt]$\phi^a$}] (1) at (-4, 2.5) {};
				\node [style=none,label={[xshift=10pt,yshift=-7pt]$\phi^b$} ] (4) at (-4, 1) {};
				\node [style=none,label={[xshift=10pt,yshift=-14pt]$\phi^c$}] (2) at (-4, -0.5) {};
			\end{pgfonlayer}
			\begin{pgfonlayer}{edgelayer}
				\draw [style=scalar] (2) to (0.center); 
				\draw [style=scalar] (1) to (0.center);
				\draw [style=scalar] (0) to (4.center);
			\end{pgfonlayer}
		\end{tikzpicture}
		& = - \frac{i}{2} \cdot \{ [ \underbrace{\Tr( T^a \bar T^b T^c)}_{\tikz{\pic[scale=0.3]{O2_ex}}}  + \text{ perms.}] + \hc\},
		\\
		\begin{tikzpicture}
			\begin{pgfonlayer}{nodelayer}
				\node [style=Op] (0) at (-6, 1) {};
				\node [style=none,label={[xshift=5pt,yshift=-1pt]$O_3$}] (3) at (-7, 1) {};
				\node [style=none,label={[xshift=10pt,yshift=0pt]$\phi^a$}] (1) at (-4, 2.5) {};
				\node [style=none,label={[xshift=10pt,yshift=-7pt]$\phi^b$} ] (4) at (-4, 1) {};
				\node [style=none,label={[xshift=10pt,yshift=-14pt]$\phi^c$}] (2) at (-4, -0.5) {};
			\end{pgfonlayer}
			\begin{pgfonlayer}{edgelayer}
				\draw [style=scalar] (2) to (0.center); 
				\draw [style=scalar] (1) to (0.center);
				\draw [style=scalar] (0) to (4.center);
			\end{pgfonlayer}
		\end{tikzpicture}
		& = - \frac{i}{2} \cdot \{ [ \underbrace{\Tr( T^a \bar T^b ) \Tr (T^c)}_{\tikz{\pic[scale=0.3]{O5_ex}}}  + \text{ perms.}] + \hc\}.
		\\
		\begin{tikzpicture}
			\begin{pgfonlayer}{nodelayer}
				\node [style=Op] (0) at (-6, 1) {};
				\node [style=none,label={[xshift=5pt,yshift=-1pt]$O_4$}] (3) at (-7, 1) {};
				\node [style=none,label={[xshift=10pt,yshift=0pt]$\phi^a$}] (1) at (-4, 2.5) {};
				\node [style=none,label={[xshift=10pt,yshift=-7pt]$\phi^b$} ] (4) at (-4, 1) {};
				\node [style=none,label={[xshift=10pt,yshift=-14pt]$\phi^c$}] (2) at (-4, -0.5) {};
			\end{pgfonlayer}
			\begin{pgfonlayer}{edgelayer}
				\draw [style=scalar] (2) to (0.center); 
				\draw [style=scalar] (1) to (0.center);
				\draw [style=scalar] (0) to (4.center);
			\end{pgfonlayer}
		\end{tikzpicture}
		& = - \frac{i}{2} \cdot \{ [ \underbrace{\Tr( T^a  T^b T^c)}_{\tikz{\pic[scale=0.3]{O1_ex}}}  + \text{ perms.}] + \hc\},
	\end{align*}
	\caption{Feynman rules for dimension-three operators.
		All the operators break the flavor symmetry $G$ by ``flipping'' the ``chirality'' of the flavor flow, which in the case of scalar operators we indicate by a box with a cross inside.  
	}
	\label{fig:O3_feynman_rules}
\end{figure*}
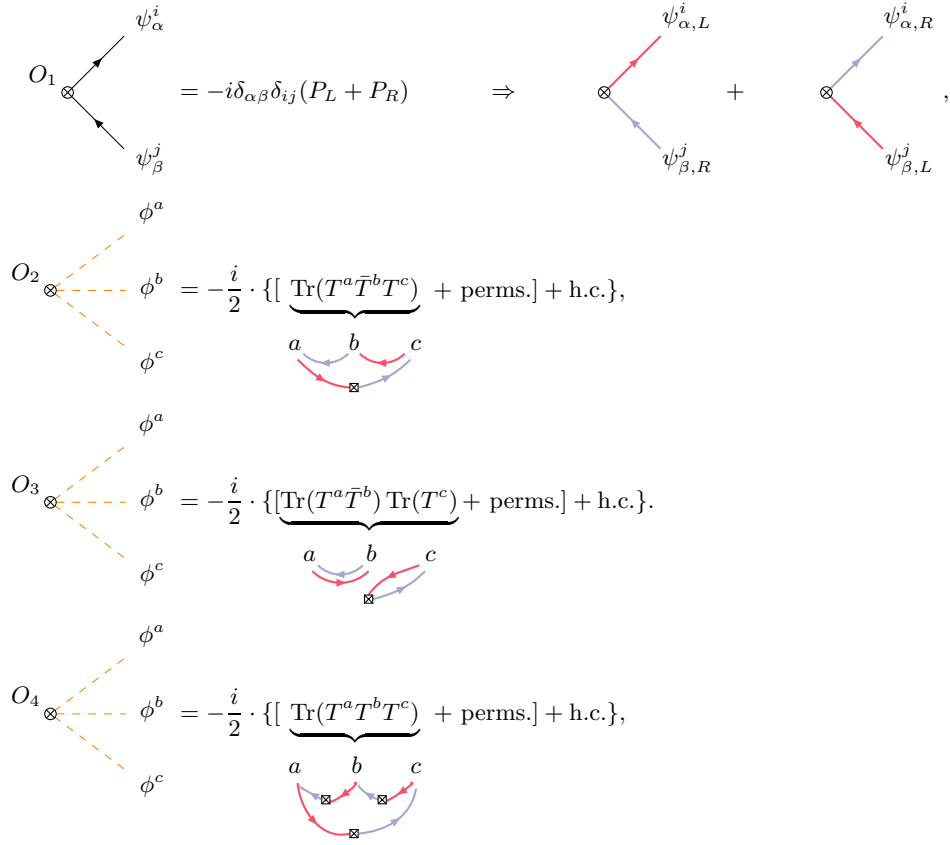

\begin{figure*}[htbp]
	\begin{align*}
		\begin{tikzpicture}
			\begin{pgfonlayer}{nodelayer}
				\node [style=Op] (0) at (-6, 1) {};
				\node [style=none,label={[xshift=5pt,yshift=-1pt]$O_5$}] (3) at (-7, 1) {};
				\node [style=none,label={[xshift=10pt,yshift=0pt]$\phi^a$}] (1) at (-4, 2.5) {};
				\node [style=none,label={[xshift=10pt,yshift=-7pt]$\phi^b$} ] (4) at (-4, 1) {};
				\node [style=none,label={[xshift=10pt,yshift=-14pt]$\phi^c$}] (2) at (-4, -0.5) {};
			\end{pgfonlayer}
			\begin{pgfonlayer}{edgelayer}
				\draw [style=scalar] (2) to (0.center); 
				\draw [style=scalar] (1) to (0.center);
				\draw [style=scalar] (0) to (4.center);
			\end{pgfonlayer}
		\end{tikzpicture}
		& = - \frac{i}{2} \cdot \{ [ \underbrace{\Tr( T^a  T^b ) \Tr (T^c)}_{\tikz{\pic[scale=0.3]{O3_ex}}}  + \text{ perms.}] + \hc\}.
		\\
		\begin{tikzpicture}
			\begin{pgfonlayer}{nodelayer}
				\node [style=Op] (0) at (-6, 1) {};
				\node [style=none,label={[xshift=5pt,yshift=-1pt]$O_6$}] (3) at (-7, 1) {};
				\node [style=none,label={[xshift=10pt,yshift=0pt]$\phi^a$}] (1) at (-4, 2.5) {};
				\node [style=none,label={[xshift=10pt,yshift=-7pt]$\phi^b$} ] (4) at (-4, 1) {};
				\node [style=none,label={[xshift=10pt,yshift=-14pt]$\phi^c$}] (2) at (-4, -0.5) {};
			\end{pgfonlayer}
			\begin{pgfonlayer}{edgelayer}
				\draw [style=scalar] (2) to (0.center); 
				\draw [style=scalar] (1) to (0.center);
				\draw [style=scalar] (0) to (4.center);
			\end{pgfonlayer}
		\end{tikzpicture}
		& = - \frac{i}{2} \cdot \{ [ \underbrace{\Tr( T^a  T^b ) \Tr (\bar T^c)}_{\tikz{\pic[scale=0.3]{O4_ex}}}  + \text{ perms.}] + \hc\}.
		\\
		\begin{tikzpicture}
			\begin{pgfonlayer}{nodelayer}
				\node [style=Op] (0) at (-6, 1) {};
				\node [style=none,label={[xshift=5pt,yshift=-1pt]$O_7$}] (3) at (-7, 1) {};
				\node [style=none,label={[xshift=10pt,yshift=0pt]$\phi^a$}] (1) at (-4, 2.5) {};
				\node [style=none,label={[xshift=10pt,yshift=-7pt]$\phi^b$} ] (4) at (-4, 1) {};
				\node [style=none,label={[xshift=10pt,yshift=-14pt]$\phi^c$}] (2) at (-4, -0.5) {};
			\end{pgfonlayer}
			\begin{pgfonlayer}{edgelayer}
				\draw [style=scalar] (2) to (0.center); 
				\draw [style=scalar] (1) to (0.center);
				\draw [style=scalar] (0) to (4.center);
			\end{pgfonlayer}
		\end{tikzpicture}
		& = - \frac{i}{2} \cdot \{ [ \underbrace{\Tr( T^a)\Tr( T^b ) \Tr (T^c)}_{\tikz{\pic[scale=0.3]{O7_ex}}}  + \text{ perms.}] + \hc\}.
		\\
		\begin{tikzpicture}
			\begin{pgfonlayer}{nodelayer}
				\node [style=Op] (0) at (-6, 1) {};
				\node [style=none,label={[xshift=5pt,yshift=-1pt]$O_8$}] (3) at (-7, 1) {};
				\node [style=none,label={[xshift=10pt,yshift=0pt]$\phi^a$}] (1) at (-4, 2.5) {};
				\node [style=none,label={[xshift=10pt,yshift=-7pt]$\phi^b$} ] (4) at (-4, 1) {};
				\node [style=none,label={[xshift=10pt,yshift=-14pt]$\phi^c$}] (2) at (-4, -0.5) {};
			\end{pgfonlayer}
			\begin{pgfonlayer}{edgelayer}
				\draw [style=scalar] (2) to (0.center); 
				\draw [style=scalar] (1) to (0.center);
				\draw [style=scalar] (0) to (4.center);
			\end{pgfonlayer}
		\end{tikzpicture}
		& = - \frac{i}{2} \cdot \{ [ \underbrace{\Tr( T^a)\Tr( T^b ) \Tr (\bar T^c)}_{\tikz{\pic[scale=0.3]{O8_ex}}}  + \text{ perms.}] + \hc\}.
	\end{align*}
	\caption{Feynman rules for dimension-three operators.
		All the operators break the flavor symmetry $G$ by ``flipping'' the ``chirality'' of the flavor flow, which in the case of scalar operators we indicate by a box with a cross inside.  
	}
	\label{fig:O3_feynman_rules_2}
\end{figure*}

In Fig.~\ref{fig:O4_feynman_rules_single}-\ref{fig:O4_feynman_rules_trip_quadr}, we also indicate the flavor ``flow'' for the single-, double-, triple- and quadruple-trace scalar couplings that can be associated with the left ($L$) and right ($R$) chiral fermions. In the absence of $U_L(N_f) \times U_R(N_f)$ breaking, the ``left-'' and ``right-handed'' flows are ``conserved'' separately.

\begin{figure*}[htbp]
	\begin{align*}
		\begin{tikzpicture}
			\begin{pgfonlayer}{nodelayer}
				\node [style=ffh] (0) at (-6, 1) {};
				\node [style=none,label={[xshift=1pt,yshift=-2pt]$\psi_{\alpha,L}^{i}$}] (1) at (-7.5, 2.5) {};
				\node [style=none,label={[xshift=-5pt,yshift=-14pt]$\psi_{\beta,R}^{j}$}] (2) at (-7.5, -0.5) {};
				\node [style=none,label={[xshift=8pt,yshift=-1pt]$\phi^a$}] (3) at (-4.5, 1) {};
				\node [style=none,minimum size=4pt,label={[xshift=-1pt,yshift=-4pt]$$}] (a) at (-4.5, 0.7) {};
				\node [style=none,minimum size=4pt,label={}] (b) at (-6.5,-0.4) {};
				\node [style=none,minimum size=4pt,label={[xshift=-1pt,yshift=-4pt]$$}] (c) at (-4.5, 1.3) {};
				\node [style=none,minimum size=4pt,label={[xshift=1pt,yshift=-4pt]$$}] (d) at (-6.5,2.4) {};
			\end{pgfonlayer}
			\begin{pgfonlayer}{edgelayer}
				\draw [style=scalar] (0) to (3.center);
				\draw [style=fermion,PLL] (2.center) to (0);
				\draw [style=fermion,PLL] (0) to (1.center);
				\draw [style={fermion},PR] (b.west) to[out=50,in=180] (a.west);
				\draw [style={fermion},PL] (c.west) to[out=-180,in=-50] (d.west);
			\end{pgfonlayer}
		\end{tikzpicture}
		= - i y_{1} T^a_{ij} \delta_{\alpha\beta}
		\hspace{4.5cm}
		\begin{tikzpicture}
			\begin{pgfonlayer}{nodelayer}
				\node [style=ffh] (0) at (-6, 1) {};
				\node [style=ChFlip] (8) at (-5.75, 1.55) {};
				\node [style=ChFlip] (9) at (-5.75, 0.45) {};
				\node [style=none,label={[xshift=1pt,yshift=-2pt]$\psi_{\alpha,L}^{i}$}] (1) at (-7.5, 2.5) {};
				\node [style=none,label={[xshift=-5pt,yshift=-14pt]$\psi_{\beta,R}^{j}$}] (2) at (-7.5, -0.5) {};
				\node [style=none,label={[xshift=8pt,yshift=-1pt]$\phi^a$}] (3) at (-4.5, 1) {};
				\node [style=none,minimum size=4pt,label={[xshift=-1pt,yshift=-4pt]$$}] (a) at (-4.5, 0.7) {};
				\node [style=none,minimum size=4pt,label={}] (b) at (-6.5,-0.4) {};
				\node [style=none,minimum size=4pt,label={[xshift=-1pt,yshift=-4pt]$$}] (c) at (-4.5, 1.3) {};
				\node [style=none,minimum size=4pt,label={[xshift=1pt,yshift=-4pt]$$}] (d) at (-6.5,2.4) {};
			\end{pgfonlayer}
			\begin{pgfonlayer}{edgelayer}
				\draw [style=scalar] (0) to (3.center);
				\draw [style=fermion,PLL] (2.center) to (0);
				\draw [style=fermion,PLL] (0) to (1.center);
				\draw [style={fermion},PR] (b.west) to[out=50,in=-140] (9.west);
				\draw [style={fermion},PL] (9.west) to[out=20,in=180] (a.west);
				\draw [style={fermion},PR] (c.west) to[out=-180,in=-20] (8.center);
				\draw [style={fermion},PL] (8.center) to[out=160,in=-50] (d.west);
			\end{pgfonlayer}
			\end{tikzpicture}
		& = 
		- i y_{2}   \bar T^a_{ij}  \delta_{\alpha \beta},
	\end{align*}
	\caption{Feynman rules for Yukawa vertices.
		The indices $i,j=1,\ldots,N_f$ count fermion generations (flavor), while $\alpha,\beta=1,\ldots,N_c$ correspond to the $SU(N_c)$ gauge group. Flavor flows for left- and right-handed fermions are indicated.
	}
	\label{fig:O4_feynman_rules_1}
\end{figure*} 	
\begin{figure*}[htbp]
	\begin{align*}
		\begin{tikzpicture}
			\begin{pgfonlayer}{nodelayer}
				\node [style=h4] (0) at (-6, 1) {};
				\node [style=none,label={[xshift=-10pt,yshift=-2pt]$\phi^a$}] (5) at (-7.5, 2.5) {};
				\node [style=none,label={[xshift=10pt,yshift=-2pt]$\phi^b$}] (2) at (-4.5, 2.5) {};
				\node [style=none,label={[xshift=-10pt,yshift=-12pt]$\phi^d$}] (3) at (-7.5, -0.5) {};
				\node [style=none,label={[xshift=10pt,yshift=-12pt]$\phi^c$}] (4) at (-4.5, -0.5) {};
			\end{pgfonlayer}
			\begin{pgfonlayer}{edgelayer}
				\draw [style=scalar] (5.center) to (0);
				\draw [style=scalar] (2.center) to (0);
				\draw [style=scalar] (0) to (4.center);
				\draw [style=scalar] (0) to (3.center);
			\end{pgfonlayer}
		\end{tikzpicture}
		& = 
		- i 2 \{
		+u \cdot [ \underbrace{\Tr (T^a \bar T^b T^c \bar T^d)}_{\tikz{\pic[scale=0.3]{v-vertex}}} + \text{ perms.}]
		+ s_1 \cdot [ \underbrace{\Tr (T^a  T^b T^c  T^d)}_{\tikz{\pic[scale=0.3]{s1}}} + \text{perms.}]\nonumber\\
		&
		+ s_2 \cdot [ \underbrace{\Tr (T^a  T^b \bar T^c  \bar T^d)}_{\tikz{\pic[scale=0.3]{s2}}} + \text{perms.}]
		+ s_3 \cdot [ \underbrace{\Tr (T^a  T^b T^c  \bar T^d)}_{\tikz{\pic[scale=0.3]{s3}}} + \text{perms.}]\}
		\nonumber\\
	\end{align*}
	\caption{Feynman rules for single trace quartic vertices. 	}
	\label{fig:O4_feynman_rules_single}
\end{figure*}

\begin{figure*}[htbp]
	\begin{align*}
		\begin{tikzpicture}
			\begin{pgfonlayer}{nodelayer}
				\node [style=h4] (0) at (-6, 1) {};
				\node [style=none,label={[xshift=-10pt,yshift=-2pt]$\phi^a$}] (5) at (-7.5, 2.5) {};
				\node [style=none,label={[xshift=10pt,yshift=-2pt]$\phi^b$}] (2) at (-4.5, 2.5) {};
				\node [style=none,label={[xshift=-10pt,yshift=-12pt]$\phi^d$}] (3) at (-7.5, -0.5) {};
				\node [style=none,label={[xshift=10pt,yshift=-12pt]$\phi^c$}] (4) at (-4.5, -0.5) {};
			\end{pgfonlayer}
			\begin{pgfonlayer}{edgelayer}
				\draw [style=scalar] (5.center) to (0);
				\draw [style=scalar] (2.center) to (0);
				\draw [style=scalar] (0) to (4.center);
				\draw [style=scalar] (0) to (3.center);
			\end{pgfonlayer}
		\end{tikzpicture}
		& = 
		- i 2 \{+ v \cdot [ \underbrace{\Tr(T^a \bar T^b) \Tr(T^c \bar T^d)}_{\tikz{\pic[scale=0.3]{u-vertex}}} + \text{perms.}]
		+ d_1 \cdot [ \underbrace{\Tr(T^a  T^b) \Tr(\bar T^c \bar T^d)}_{\tikz{\pic[scale=0.3]{d1}}} + \text{perms.}]
		\nonumber\\
		& +  d_2 \cdot [ \underbrace{\Tr(T^a  T^b) \Tr( T^c  T^d)}_{\tikz{\pic[scale=0.3]{d2}}} + \text{ perms.}]
		+  d_3 \cdot [ \underbrace{\Tr(T^a  T^b) \Tr( T^c \bar T^d)}_{\tikz{\pic[scale=0.3]{d3}}} + \text{ perms.}]
		\nonumber\\
		& +  d_4 \cdot [ \underbrace{\Tr(T^a  T^b T^c) \Tr(   T^d)}_{\tikz{\pic[scale=0.3]{d4}}} + \text{ perms.}]
		+  d_5 \cdot [ \underbrace{\Tr(T^a   T^b T^c) \Tr(  \bar T^d)}_{\tikz{\pic[scale=0.3]{d5}}} + \text{ perms.}]
		\nonumber\\
		& +  d_6 \cdot [ \underbrace{\Tr(T^a \bar T^b T^c) \Tr(   T^d)}_{\tikz{\pic[scale=0.3]{d6}}} + \text{ perms.}]
		+  d_7 \cdot [ \underbrace{\Tr(T^a  \bar T^b T^c) \Tr(  \bar T^d)}_{\tikz{\pic[scale=0.3]{d7}}} + \text{ perms.}]
		\}
	\end{align*}
	\caption{Feynman rules for double trace quartic vertices. 	}
	\label{fig:O4_feynman_rules_double}
\end{figure*}

\begin{figure*}[htbp]
	\begin{align*}
		\begin{tikzpicture}
			\begin{pgfonlayer}{nodelayer}
				\node [style=h4] (0) at (-6, 1) {};
				\node [style=none,label={[xshift=-10pt,yshift=-2pt]$\phi^a$}] (5) at (-7.5, 2.5) {};
				\node [style=none,label={[xshift=10pt,yshift=-2pt]$\phi^b$}] (2) at (-4.5, 2.5) {};
				\node [style=none,label={[xshift=-10pt,yshift=-12pt]$\phi^d$}] (3) at (-7.5, -0.5) {};
				\node [style=none,label={[xshift=10pt,yshift=-12pt]$\phi^c$}] (4) at (-4.5, -0.5) {};
			\end{pgfonlayer}
			\begin{pgfonlayer}{edgelayer}
				\draw [style=scalar] (5.center) to (0);
				\draw [style=scalar] (2.center) to (0);
				\draw [style=scalar] (0) to (4.center);
				\draw [style=scalar] (0) to (3.center);
			\end{pgfonlayer}
		\end{tikzpicture}
		& = 
		& - i 2 \{
		+  t_1 \cdot [ \underbrace{\Tr(T^a  T^b ) \Tr(T^c)\Tr(T^d)}_{\tikz{\pic[scale=0.3]{t1}}} + \text{ perms.}]{}\nonumber\\ 
		& \hspace{0.5cm}
		&+  q_1 \cdot [ \underbrace{\Tr(T^a) \Tr(T^b)\Tr(T^c)\Tr(T^d)}_{\tikz{\pic[scale=0.3]{q1}}} + \text{ perms.}]
		\}
	\end{align*}
	\caption{Feynman rules for triple and quadruple trace quartic vertices. 	The same diagrams arise for other operators but with changed right- and left-handed flows.}
	\label{fig:O4_feynman_rules_trip_quadr}
\end{figure*}

\section*{Acknowledgement}
A.I.M. would like to thank Alexander Bednyakov for fruitful discussions at various stages of the work, and Tom Steudtner and Daniel Litim  for helpful suggestions.  
Some of the results were presented at the 12th International Conference on the Exact Renormalization Group, Les Diablerets (Sept 2024).  
The work is supported by the Foundation for the Advancement of Theoretical Physics and Mathematics BASIS, No 24-1-4-36-1.

\bibliography{biblio.bib}
\end{document}